\documentclass[a4paper,11pt]{article}
\usepackage{jcappub} % for details on the use of the package, please see the JCAP-author-manual
\usepackage{verbatim}
\usepackage{multirow}
\usepackage{float}
\restylefloat{table}
\usepackage{pifont}
\usepackage{epsfig}
\usepackage{graphicx}
\usepackage{boldline}
\usepackage{booktabs}
\usepackage{multirow}
\usepackage{dcolumn}
\usepackage{amsmath}
\usepackage{mathtools}
\usepackage{amsfonts}
\usepackage{amssymb}
\usepackage{epstopdf}
\usepackage{multirow,bigdelim}
\usepackage{bm}
\usepackage{braket}
\usepackage{enumitem}
\usepackage{soul}
\usepackage[table]{xcolor}
\definecolor{navyblue}{rgb}{0.0, 0.0, 0.5}
\definecolor{royalblue}{rgb}{0.25, 0.41, 0.88}
\definecolor{cadmiumgreen}{rgb}{0.0, 0.42, 0.24}
\definecolor{blue-violet}{rgb}{0.54, 0.17, 0.89}
\definecolor{darkviolet}{rgb}{0.58, 0.0, 0.83}
\definecolor{orange(colorwheel)}{rgb}{1.0, 0.5, 0.0}

\usepackage{hyperref}
\hypersetup{colorlinks=true,linkcolor=royalblue,citecolor=magenta,pdfencoding=auto}
\usepackage{pifont}
\usepackage{dcolumn}
\usepackage{colortbl}

\newcommand{\ba}{\begin{eqnarray}}
\newcommand{\ea}{\end{eqnarray}}
\newcommand{\ep}{\epsilon}
\newcommand\be{\begin{eqnarray}}
\newcommand\ee{\end{eqnarray}}

\newcommand{\R}{{\cal{R}}}

\newcommand\eea{\end{eqnarray}}
\newcommand\bea{\begin{eqnarray}}

\newcommand{\nn}{\nonumber}

\newcommand{\bp}{{\bf p}}

\newcommand{\bx}{{\bf x}}

\newcommand{\bfx}{{\bf{x}}}

\def\bc{{\bf c}}

\def\bp{{\bf p}}

\def\bx{{\bf x}}

\def \ep {\epsilon}

\def \bc {C}
\def \bs{\Sigma}
\def \bom{\Omega}
\def\then{\quad \Rightarrow\quad}
\def\refeq{\eqref}
\def\Rc{\mathcal{R}_{c}}
\def\No{N_{1}}

\def \ep {\epsilon}
\def \bzeta {\bar{\zeta}}
\def \baeta{\bar{\eta}}
\def \brho{\bar{\rho}}
\def \bp {\bar{p}}

\renewcommand\[{\left[}

\newcommand{\e}{\epsilon}

\definecolor{magenta(process)}{rgb}{1.0, 0.0, 0.56}

\definecolor{darkspringgreen}{rgb}{0.09, 0.45, 0.27}

\definecolor{royalblue(web)}{rgb}{0.25, 0.41, 0.88}

\title{\centering Systematics of Adiabatic Modes:\\ Flat Universes}

\author[a]{E. Pajer,}
\author[a,b]{S. Jazayeri}%\note{Also at Some University.}}

\affiliation[a]{Institute for Theoretical Physics and Center for Extreme Matter and Emergent Phenomena,
	Utrecht University, %\\ 
	Princetonplein 5, 3584 CC Utrecht, The Netherlands}
\affiliation[b]{School of Astronomy, Institute for Research in Fundamental Sciences (IPM) \\ P.~O.~Box 19395-5531, Tehran, Iran }
\emailAdd{e.pajer@uu.nl}
\emailAdd{sadraj@ipm.ir}

\abstract{\noindent Adiabatic modes are cosmological perturbations that are locally indistinguishable from a (large) change of coordinates. At the classical level, they provide model independent solutions. At the quantum level, they lead to soft theorems for cosmological correlators. We present a systematic derivation of adiabatic modes in spatially-flat cosmological backgrounds with asymptotically-perfect fluids. We find several new adiabatic modes including vector, time-dependent tensor and time-dependent scalar modes. The new vector and tensor modes decay with time in standard cosmologies but are the leading modes in contracting universes. We present a preliminary derivation of the related soft theorems. In passing, we discuss a distinction between classical and quantum adiabatic modes, we clarify the subtle nature of Weinberg's second adiabatic mode and point out that the adiabatic nature of a perturbation is a gauge dependent statement.}

\begin{document}
\maketitle
\flushbottom

\vspace{1cm}
\begin{flushright}
\textit{``Heard about the guy who fell off a skyscraper? \\On his way down past each floor, he kept saying to reassure himself: So far so good... so far so good... so far so good. How you fall doesn't matter. It's how you land!''\\[0.3cm] M. Kassovitz, La Haine (1995)}
\end{flushright}

\section{Introduction}

In particle physics, as we descend from the highest energies that are generated in particle colliders, we cannot help feeling safe and reassured. Unknown, irrelevant operators disappear into tiny corrections. The ranks of subatomic particles shrink steadily as high-mass excitations become forbidden by energy conservation. So far so good. We do encounter some bumps along the road, such as the strong coupling of Quantum Chromodynamics, but we move right through it thanks to spontaneously broken symmetries and (pseudo-) Goldstone Bosons. So far so good. At very low energies, by particle physics standards, we enjoy the comforting 5-loop verification of Quantum Electrodynamics. So far so good. But as we descend further in energy, at distances much larger than particle detectors, dark shadows enter the scene. Already on galactic scales, our cherished laws of gravitation oblige us to postulate the existence of a new form of (Dark) matter. At the largest possible distances theoretically accessible, we observe the accelerated expansion of the universe and we are forced to postulate the existence of some yet unknown and unconventional form of (Dark) energy. Moreover, the distribution of everything on cosmological scales is at odds with causality in the old Hot Big Bang model (horizon problem). Yet again, this forces us to invoke additional beyond-the-standard-model physics, such as the inflaton, and a specific primordial dynamics, such as inflation. To tackle these and other long-distances phenomena that we access though cosmological observations, we naturally turn our attention to the infra-red structure of gravity. In this work, we make progress in this direction by presenting several infinite classes of new non-linear symmetries of perturbations around cosmological backgrounds and the related model-independent linear solutions, which extend the known cosmological adiabatic modes \cite{WAD,PaoloCF,Khoury1}. Each symmetry in turn generates soft theorems that must be obeyed by cosmological correlators.\\

In \cite{WAD} Weinberg introduced cosmological adiabatic modes\footnote{They are constructed as follows. After imposing a local gauge condition that fixes all finite momentum diffs, there are still residual large diffeomorphism that obey the gauge choice. Some of these residual diffs can be extended to finite momentum and become adiabatic modes.} as physical, finite momentum perturbations around an FLRW background that, in the zero momentum limit, become arbitrarily close to some large gauge transformation (a diffeomorphism that does not vanish at spatial infinity).  Each adiabatic mode can be related to a non-linear symmetry of the action for perturbations \cite{Khoury1,HinterbiCHLer:2013dpa}. Adiabatic modes can be thought of as Goldstone Bosons that non-linearly realize large residual diffs. They differ from particle physics Goldstone bosons in a few respects. First, the lack of a time-like Killing vector of the background makes it hard to define what we mean by gapless excitations. Second, the residual large diffs lead to non-uniform symmetries, which depend explicitly on spacetime (as opposed to just through the fields). This is a serious obstacle to generalize the well-known Goldstone Boson properties such as decoupling at zero-momentum. Third, all diffs are gauged by gravity and adiabatic modes turn out to be gauge dependent. \\

Knowing the existence of an adiabatic mode is useful for several reasons. First, it tells us about the existence of a linear solution that is model independent (up to some technical assumptions on anisotropic stresses and Fourier space asymptotia). This is very useful because we do not know the matter content of the early stages of cosmic evolution, e.g. during and right after reheating. Second, adiabatic modes can be used to derive soft theorems such as Maldacena's consistency relation \cite{Maldacena:2002vr}, which provide model independent constraints on cosmic correlators. Third, the construction of adiabatic modes highlights the fact that certain correlations are artifacts of the gauge choice and are not locally measurable. In this sense, adiabatic modes help identify local observables that are perturbatively gauge-invariant, in an equivalent but independent way from the use of Fermi Normal Coordinates and their Conformal generalization \cite{Pajer:2013ana,Dai:2015jaa,Dai:2015rda}. Fourth, adiabatic modes might provide the right language to connect to recent developments in the infra-red structure of gauge and gravity theories around Minkowski spacetime \cite{Strominger:2017zoo}, along the lines of \cite{Mirbabayi:2016xvc,AntonioRiotto}. Finally, in stark contrast with flat space amplitudes, it is unclear what the general, model independent rules are to write down consistent cosmological correlators. Soft theorems for adiabatic modes provide at least some robust checks that cosmological correlators must satisfy.\\

There are at least two reasons to suspect that more adiabatic modes should exist beside those discovered in \cite{PaoloCF,WAD,HinterbiCHLer:2013dpa}. First, scalar and tensor adiabatic modes are known, but vector adiabatic modes are surprisingly absent. Second, tensor adiabatic modes in the literature correspond only to one of the two linear solutions of the equations of motion, the ``time independent'' mode. This is suspicious since, in the absence of matter, gravitational waves should be locally a gauge artifact and so one suspects both modes should be adiabatic. \\

In the following we summarize the new adiabatic modes we found around flat FLRW spacetimes with a single fluid that becomes perfect on large scales. The generalization to multiple fluids is straightforward.
\begin{itemize}
\item We present \textit{vector adiabatic modes} to all orders in spatial gradients and discuss them in detail in Newtonian and comoving gauge. As expected, they decay in time in standard cosmologies.
\item We find a \textit{new scalar adiabatic mode} corresponding to the decaying mode of curvature perturbations. Interestingly, this mode exists for a generic perfect fluid, but it disappears for a single generic scalar field\footnote{The new scalar adiabatic mode survives only if the scalar field is shift symmetric and leads to soft theorems for Ultra Slow Roll inflation \cite{toappear}.}.
\item We show that also the \textit{time-dependent solution of tensor modes is adiabatic}, when mixed with higher order gradients in scalars or vectors. The new mode decays in standard cosmology but is the leading one in contracting universes. 
\item We present preliminary results for the soft theorems related to vector and tensor adiabatic modes. These might be useful to characterize alternatives to inflation that invoke a primordial phase of contraction.
\end{itemize}

The rest of the paper is organized as follows. We systematically construct all adiabatic modes for a general single fluid around a flat FLRW background in Newtonian gauge in Section \ref{ss:Newtonian}, and in comoving gauge in Section \ref{ssec:comoving}. We discuss in details the new adiabatic modes we found and their properties in Section \ref{ssec:prop}. In Section \ref{ssec:soft}, we present a preliminary discussion of the related soft theorems. Finally, we conclude in Section \ref{concl}. The appendices contain a series of explicit results and computational details.
% 
%
%\paragraph{Notation and conventions} We use mostly plus signature and raise and lower time and spatial indices $  i, j, \dots $ with the background spatial metric $  \bar g_{\mu\nu} $. 

%%%%%%%%%%%%%%%%%%%%%%%%%%%%%%%%%%%%%%%%

\section{Newtonian Gauge}\label{ss:Newtonian}

To introduce our notation\footnote{We use the notation of \cite{WB}.}, let us recall the general scalar-vector-tensor decomposition of metric perturbations in linear cosmological perturbation theory around flat FLRW spacetimes
\ba
ds^2=-(1+E)dt^2+2a (\partial_i F+G_i)dt dx^i+a^2 \left[ (1+A)\delta_{ij}+\partial_{i}\partial_{j}B+2\partial_{(i} C_{j)}+D_{ij} \right]\,,
\ea
where
\ba
D_{ii}=\partial_i D_{ij}=\partial_i C_i=\partial_i G_i&=&0 \,.
\ea
For simplicity, we focus on a system with a single perfect fluid
\be
T_{\mu\nu}=\left(  \rho+p\right)u_{\mu}u_{\nu}+pg_{\mu\nu}\,,
\ee
with $  u_{\mu}u^{\mu}=-1 $. All of our results can be straightforwardly generalized to the multiple fluids by substituting $\delta u$ and $\delta\rho/\dot{\rho}$ with a sum of $\delta u_a$ and $ \delta \rho_a /\dot{\rho}_a$ over each fluid, respectively. Also if we assume that the fluid anisotropic stress vanishes on super-Hubble scales, our results generalize to non-perfect fluids as well. Our scalar-vector-tensor decomposition of the fluid velocity is
\be
u_\mu=(u_0,u_i)\,, \quad u_i=\partial_i \delta u+\delta u^V_i\,, \quad \partial_i\delta u_i^V=0\,.
\ee
We use the following definitions of (small-)gauge-invariant curvature perturbations on comoving and homogeneous slices, respectively 
\ba
{\cal R}&=& \dfrac{A}{2}+H\delta u\,,\\ 
{\cal \zeta}&=&\dfrac{A}{2}-H\dfrac{\delta \rho}{\dot{\rho}}\,.
\ea
In this section, we specialize to Newtonian gauge, while comoving gauge is discussed in \ref{ssec:comoving}.

By a gauge transformation (choice of coordinates) we can remove two scalars and one vector mode from the metric. The Newtonian gauge condition is
\be
B=F=0 \quad \text{and} \quad C_i=0\,.
\ee
Therefore, the form of the Newtonian gauge is
\begin{equation}
\label{NG}
ds^2=-(1+2\Phi)dt^2+a(t)G_i dt dx^i+a(t)^2(1-2\Psi)\delta_{ij}dx^idx^j+a(t)^2 \gamma_{ij}dx^idx^j\,,
\end{equation}
where, to follow the usual notation, we renamed $  E=2\Phi $, $  A=-2\Psi $ and $  D_{ij}=\gamma_{ij} $. Here, $  \gamma_{ij} $ is a transverse traceless two-tensor, $\partial_{i}\gamma_{ij}=\gamma_{ii}=0 $. Under an arbitrary, infinitesimal diffeomorphism
\be
x^{\mu}\rightarrow x^{\mu}+\epsilon^{\mu}(x)\,,
\ee 
the different components of the metric transform as
\bea 
\label{trans}
\Delta h_{ij}&=&2a^2H\ep_0\delta_{ij}-\bar g_{ik}\partial_j \ep^k-\bar g_{jk}\partial_i \ep^k \,,\\ \nn
&=&2a^2H\delta_{ij}\epsilon_0-2\ep_{(i,j)}\label{ij}\,,\\ 
\Delta h_{0i}&=&-\dot{\ep}_i-\partial_i \epsilon_0+2H\ep_i \,,\label{trans2}\\ 
\Delta h_{00}&=&-2\dot{\ep}_0\,,\label{trans3}
\eea
at linear order in metric perturbations, $  g_{\mu\nu}\equiv \bar g_{\mu\nu}+h_{\mu\nu} $. The changes in the matter fields are given by 
\ba
\frac{\Delta \delta \rho}{\dot{\rho}}=\frac{\Delta \delta p}{\dot{p}} &=& \epsilon_0\,,\\ 
\Delta \left(\delta u^{V}_{i}+\partial_i \delta u\right)&=&-\partial_i \epsilon_0\,.
\ea
Note that we raise and lower spacetime indices with the background metric, $  \bar g_{00}=-1$ and $\bar g_{ij}=a^{2}\delta_{ij} $, so
\be
\e^{i}=\dfrac{1}{a^2}\ep_i \quad  \text{and}\quad\epsilon_{0}=-\epsilon^{0}\,.
 \ee
To ensure that the transformed metric remains in the Newtonian gauge \eqref{NG}, $  \epsilon^{\mu} $ must satisfy the following conditions\footnote{In our notation, the symmetric and antisymmetric parts of a matrix are defined as $M_{(ij)}\equiv\dfrac{1}{2}\left[ M_{ij}+M_{ji} \right]  $, $M_{[i j]}\equiv\dfrac{1}{2}\left[ M_{ij}-M_{ji} \right]$.}
\ba
\label{epi2}
\partial_{i}\gamma_{ij}=\gamma_{ii}=0\then\nabla^2 \ep_i&=&-\dfrac{1}{3}\partial_i \partial_k \ep_k \,,\\ 
\label{ep0}
\partial_{i}h_{0i}=0\then \nabla^2\epsilon_0&=&2H\partial_i\epsilon_i-\partial_i\dot{\epsilon}_i\,. 
\ea
For \textit{small gauge transformations} $  \epsilon^{\mu} $, which vanish at spatial infinity, these conditions have no solution, reflecting the fact that the Newtonian gauge fully fixes the metric. For \textit{large gauge transformation} $  \epsilon^{\mu} $, which do not vanish at spatial infinity, $  \lim_{\bf{x}\rightarrow \infty}\epsilon^{\mu}(t,{\bf{x}})\neq 0 $, these conditions have infinitely many solutions, which we study here in detail.

Using \eqref{trans}-\eqref{trans3} one can extract the metric and matter perturbations created by the $(\epsilon_0,\epsilon_i)$ gauge transformation on an unperturbed flat FLRW universe. The result is
\ba\label{psi1}
\Psi &=& -H\epsilon_0+\dfrac{1}{3a^2}\partial_k \epsilon_k\,,\quad \Phi = \dot{\epsilon}_0\,,\quad \delta u_{i}=-\partial_i \epsilon_0\,,\\ \label{psi2}
\dfrac{\delta \rho}{\dot{\rho}}&=& \epsilon_0\,,\quad G_i =\dfrac{1}{a}\left(-\partial_i \epsilon_0+2H \epsilon_i-\dot{\epsilon}_i\right)\,,\\
\gamma_{ij}&=&\dfrac{1}{a^2}\left(-2\e_{(i,j)}+\dfrac{2}{3}\partial_k\e_k\delta_{ij}\right)
\,.\label{general}
\ea
These modes solve all Einstein equations (since they are diff-covariant) but are pure gauge and therefore unphysical. In Fourier space, they are proportional to $  \delta_{D}(\vec{q}) $ and its derivatives. However, a subset of these modes can be slightly modified to have a suitable decay at spatial infinity, or equivalently, to have support at non-vanishing $  \vec{q} $ in Fourier space. For this to be possible, one needs to check that all Einstein equations that vanish in the $q\to 0$ limit are also satisfied for small but finite $  q $  \cite{WAD}. In Newtonian gauge, the equations that vanish in the $q\to 0$ limit are the off-diagonal parts of the $ij$ components
\ba 
\label{off-s}
\partial_{i}\partial_{j}\left( \Phi-\Psi \right)&=&0\,,\\ 
\label{off-v}
\partial_{j}\left[ \dot{G}_i+2HG_i \right]&=& 0\,,
\ea
and the $0i$ components of the Einstein equations
\ba
\label{0i-s}
\dot{H}\partial_{i}\delta u &=& \partial_{i}\left( H\Phi+\dot{\Psi} \right)\,,\\
\label{0i-v} 
-4\dot{H}a \delta u^V_j &=&\nabla^2 G_j\,.
\ea
To systematically construct all adiabatic modes, we solve the above constraints and the conditions \eqref{epi2} and \eqref{ep0} in a derivative expansion\footnote{One may also postpone the derivative expansion to the last step in the derivation as in \cite{HinterbiCHLer:2013dpa}. We do this in section \ref{ssec:comoving}.}. The calculation is straightforward but tedious so we postpone it to appendix \ref{app:det} and report here only the final result. The spatial gauge parameter $\epsilon_i$ can be expanded to cubic order in $\vec{x}$
\be
\label{ei}
\e^{i}=\dfrac{1}{a^2}\ep_i\simeq c_i(t)+\omega_{il}(t)x^l+\dfrac{1}{2!}\sigma_{ikl}(t)x^kx^l+\dfrac{1}{3!}\mu_{ijkl}x^jx^kx^l\,,
\ee
where the time dependence of the coefficients is constrained to be
\bea
\sigma_{ijk}(t)&=&\tilde{\sigma}_{ijk}{\displaystyle \int \dfrac{dt'}{a(t')^3}}+\sigma^0_{ijk}\quad , \quad \tilde{\sigma}_{kki}=-3\tilde{\sigma}_{ikk}\,,\\
\mu_{ijkl}&=&\tilde{\mu}_{ijkl}{\displaystyle \int \dfrac{dt'}{a(t')^3}}+{\mu}^0_{ijkl}\quad , \quad \tilde{\mu}_{iikl}=-3\tilde{\mu}_{klii}\,,\\ \nn
\omega_{ij}(t)&=&\tilde{\omega}_{ij}{\displaystyle\int \dfrac{dt'}{a(t')^3}+\omega^0_{ij}}-\dfrac{1}{3}\mu^0_{kkij}\int^t \dfrac{dt'}{a(t')^3}\int dt''a(t'')\\
&&+\dfrac{1}{4}{\cal D}^0_{ijkk}\int^t \dfrac{dt'}{a(t')^3}\int^{t'}dt''a(t'')\int^{t''}\dfrac{dt'''}{a(t''')^3}\,,\\
c_i(t)&=&-{\displaystyle \int \dfrac{dt'}{3a(t')^3}\int^{t'} dt''a(t'')\sigma^0_{kki}+\int\dfrac{{\cal C}_i dt'}{a(t')^3}}\\ \nn
&&+\dfrac{1}{4}{\cal C}^0_{ikk}\int^t \dfrac{dt'}{a(t')^3}\int^{t'}dt''a(t'')\int^{t''}\dfrac{dt'''}{a(t''')^3}\,,
\eea
for constants $  \cal C $, $  {\cal C}_{i} $, $  \omega^{0}_{ij} $, $  \tilde \omega_{ij} $, $  \sigma_{ijk}^{0} $ and $\tilde  \sigma_{ijk} $. The analogous formula for $\epsilon_0$ is
\ba
\epsilon_0 &=&\dfrac{1}{3a}\tilde{\omega}_{ii}\int^t dt'a(t')\int^{t'}\dfrac{dt''}{a(t'')^3}+\dfrac{1}{3a}\omega^0_{ii}{\displaystyle \int dt' a(t')+\dfrac{{\cal C}}{a}}\\ \nn
&&+
\dfrac{1}{a}\left({\cal A}^0_i+\dfrac{1}{3}\sigma^0_{kki}\int dt'a(t')-{\cal C}_i-\dfrac{1}{4}{\cal C}^0_{ikk}\int^{t}dt'a(t')\int^{t'}\dfrac{dt''}{a(t'')^3}\right)x^i\\ \nn
&&+\dfrac{1}{2a}\left({\cal B}^0_{ij}-\tilde{\omega}_{ij}+\dfrac{1}{3}\mu^0_{kkij}\int^t  dt'a(t')-\dfrac{1}{4}{\cal D}^0_{ijkk}\int^t dt'a(t')\int^{t'}\dfrac{dt''}{a(t'')^3}\right)x^ix^j\\ \nn
&&+\dfrac{1}{3!a}({\cal C}^0_{ijk}-\tilde{\sigma}_{ijk})x^ix^jx^k+\dfrac{1}{4!a}\left({\cal D}^0_{ijkl}-\tilde{\mu}_{ijkl}\right)x^ix^jx^kx^l\,.
\ea
A summary of the symmetry properties of all Taylor coefficients is given in Table \ref{syms}.

\begin{table}
\begin{center}
  \begin{tabular}{|c| c |c | c | c|}
  \hline
  \rowcolor{gray}
   &definition&Constraints &No. elements \\ 
   \hline
   &&&\\
   ${\cal C}$&\ref{eppt}&-&1\\
   &&&\\
   ${\cal C}_i$ &\ref{ciit}&-& 3\\ 
   &&&\\
   ${\cal A}^0_i$&\ref{Ai}&-&3\\ 
   &&&\\
   $\omega^0_{ij}$ &\ref{omega}&symmetric & 6 \\
   &&&\\
   $\tilde{\omega}_{[ij]}$ &\ref{omega}&anti-symmetric & 3\\
   &&&\\
   $\tilde{\omega}_{(ij)}$ &\ref{omega}&symmetric & 6\\
   &&&\\
   ${\cal B}^0_{(ij)}$&\ref{Bij}& symmetric + traceless &5\\ 
   &&&\\
   ${\cal B}^0_{[ij]}$&\ref{Bij}& $=\tilde{\omega}_{[ij]}$&0\\
   &&&\\
   $\sigma^0_{ijk}$ &\ref{siggt}& sym. in j$\leftrightarrow$k + ($\sigma^0_{jji}=-3\sigma^0_{ijj}$) & 15\\ 
   &&&\\
   $\tilde{\sigma}_{ijk}$ &\ref{siggt}& sym. in j$\leftrightarrow$k + $(\tilde{\sigma}_{ikk}=-\dfrac{1}{3}\tilde{\sigma}_{kki})$ & 15\\ 
   &&&\\
  ${\cal C}^0_{ijk}$& \ref{Cijk}& 
  $\left({\cal C}^0_{ijk}-\tilde{\sigma}_{ijk}=\text{totally sym.}\right)$+$\left({\cal C}^0_{iik}=0\right)$& 7\\
  &&&\\
$\tilde{\mu}_{ijkl}$&\ref{mut}& \text{sym in jkl}+$\left(\tilde{\mu}_{ijkk}=-\dfrac{1}{3}\tilde{\mu}_{kkij}\right)$ &21\\ 
   &&&\\
   ${\cal D}^0_{ijkl}$&\ref{Dijkl}& $\left({\cal D}^0_{ijkl}-\tilde{\mu}_{ijkl}=\text{totally sym.}\right)$+$({\cal D}^0_{iikl}=0)$ & 9\\
    \hline
  \end{tabular}
\end{center}
\caption{Here we summarize the properties and number of elements of the coefficients appearing in the gradient expansion of the gauge parameter $ \e^{\mu}$. For counting the degrees of freedom, we consider ${\cal B}^0$, ${\cal C}^0$, and ${\cal D}^0$ as dependent variables, while $\tilde{\omega}$, $\tilde{\sigma}$, and $\tilde{\mu}$ are considered as independent matrices.\label{syms}}
\end{table}

%%%%%%%%%%%%%%%%%%%%%%%%%%

\begin{table}
\small
\begin{center}
\begin{tabular}{|>{\columncolor[gray]{0.92}}c|c|}
\hlineB{3}
&\\
$\epsilon_0$ &
$\begin{aligned}
&\dfrac{1}{3a}\tilde{\omega}_{ii}\int^t dt'a(t')\int^{t'}\dfrac{dt''}{a(t'')^3}+\dfrac{1}{3a}\omega^0_{ii}{\displaystyle \int dt' a(t')+\dfrac{{\cal C}}{a}}\\ 
&+
\dfrac{1}{a}\left({\cal A}^0_i+\dfrac{1}{3}\sigma^0_{kki}\int dt'a(t')-{\cal C}_i-\dfrac{1}{4}{\cal C}^0_{ikk}\int^{t}dt'a(t')\int^{t'}\dfrac{dt''}{a(t'')^3}\right)x^i\\ 
&+\dfrac{1}{2a}\left({\cal B}^0_{ij}-\tilde{\omega}_{ij}-\dfrac{1}{4}{\cal D}^0_{ijkk}\int^t dt'a(t')\int^{t'}\dfrac{dt''}{a(t'')^3}\right)x^ix^j\\
&+\dfrac{1}{3!a}({\cal C}^0_{ijk}-\tilde{\sigma}_{ijk})x^ix^jx^k+\dfrac{1}{4!a}\left({\cal D}^0_{ijkl}-\tilde{\mu}_{ijkl}\right)x^ix^jx^kx^l\,.
\end{aligned}$\\ 
&\\
\hlineB{3}
&\\ 
$\epsilon^i$ & 
$
\begin{aligned}
&-{\displaystyle \int \dfrac{dt'}{3a(t')^3}\int^{t'} dt''a(t'')\sigma^0_{kki}+\int\dfrac{{\cal C}_i dt'}{a(t')^3}}
+\dfrac{1}{4}{\cal C}^0_{ikk}\int^t \dfrac{dt'}{a(t')^3}\int^{t'}dt''a(t'')\int^{t''}\dfrac{dt'''}{a(t''')^3}\\ 
&+\left[\tilde{\omega}_{ij}{\displaystyle\int \dfrac{dt'}{a(t')^3}+\omega^0_{ij}}+\dfrac{1}{4}{\cal D}^0_{ijkk}\int^t \dfrac{dt'}{a(t')^3}\int^{t'}dt''a(t'')\int^{t''}\dfrac{dt'''}{a(t''')^3}\right]x^j\\
&+\dfrac{1}{2}\left[\tilde{\sigma}_{ijk}{\displaystyle \int \dfrac{dt'}{a(t')^3}}+\sigma^0_{ijk}\right]x^jx^k+\dfrac{1}{6}\tilde{\mu}_{ijkl}{\displaystyle \int \dfrac{dt'}{a(t')^3}}x^jx^kx^l
\end{aligned}
$\\
&\\
\hlineB{3}
\end{tabular}
\end{center}
\caption{The residual diffeomorphisms in Newtonian gauge, used for constructing the adiabatic modes, up to first order in spatial gradient.}
\end{table}

%%%%%%%%%%%%%%%%%%%%%%%%%%%%%%%%%%%%%%%%
 
\subsection{Known adiabatic modes}\label{ssec:}
 
Putting all things together, we can reproduce the known adiabatic scalar and tensor modes of \cite{WAD} and \cite{PaoloCF,Khoury1}. 
%%%%%%%%%%%%%%
\begin{itemize}
\item \textbf{Weinberg's first scalar mode} The most well-known adiabatic mode consists of a spatial dilation accompanied by a time-dependent time translation
\be
\omega_{ij}=\omega^0_{ij}=\dfrac{1}{3}\lambda \delta_{ij} \then \epsilon(t)=\dfrac{\lambda}{3a}\int dt' a(t')\,, 
\ee
where all other Taylor coefficients are set to zero. According to table \ref{diff}, this results in Weinberg's first adiabatic mode \cite{WAD}
\ba
\Phi=\Psi&=&\dfrac{\lambda}{3}\left[ 1-\dfrac{H}{a}\int dt' a(t') \right]\,, \nn \\ \label{1}
\delta u=-\dfrac{\delta \rho}{\dot{\rho}} &=& \dfrac{\lambda}{3a}\int dt' a(t')\,,\\ \nn
\zeta={\cal R}&=& -\dfrac{\lambda}{3}\,.
\ea
%%%%%%%%%%%%%%%%%%
\item \textbf{Weinberg's second scalar mode} There is a second scalar mode with spatially constant Newtonian potentials, namely
\ba\label{2}
\epsilon(t)&=&\dfrac{{\cal C}}{a}\then \Phi=\Psi=H \delta u=-\dfrac{{\cal C}H}{a}\,,\quad {\cal R}= 0\,.
\ea
This is Weinberg's second scalar adiabatic mode \cite{WAD}. In standard cosmology, namely an expanding universe, $  \dot a>0 $, that does not violate the Null Energy Condition, $  \dot H<0 $, this mode decays with time. As we will see around \refeq{2a}, this mode disappears in comoving gauge. This and other subtleties of this mode are discussed in section \ref{thirdman}.
%%%%%%%%%%%%%%%%%%%%%%%
\item \textbf{Weinberg's tensor mode} A time-independent traceless anisotropic rescaling,
\ba
\omega_{ij}= \omega^0_{ij} &,& \omega^0_{kk}=0 \then \e_{0}=0\,,
\ea
results in a spacetime constant tensor mode
\bea
\gamma_{ij}&=&\omega^0_{ij}\,.\label{3}
\eea
If the finite momentum adiabatic mode is monochromatic, then one needs to additionally impose a transversality condition on $  \omega^{0} $. While monochromaticity is useful for counting the number of independent Ward identities \cite{HinterbiCHLer:2013dpa}, as explained in \cite{MehrdadGrad} there are also adiabatic modes that are not monochromatic, i.e. they contain more than one wavevector.

%%%%%%%%%%%%%%%%%%%%%%%%%%
\item \textbf{Gradient scalars (SCT)}
This mode corresponds to a special conformal transformation (SCT) parameterized by the constant vector $  b_{i} $, accompanied by an appropriate time dependent translation
\ba
\sigma_{ijk}=\sigma^0_{ijl}&=&\delta_{il}b_j+\delta_{ij}b_l-\delta_{jl}b_i\,, \label{SCT}\\ \nn
\displaystyle c_i(t)&=& -b_i\int \dfrac{dt'}{a(t')^3}\int^{t'} dt''a(t'')\,.
\ea
The result is the gradient adiabatic mode discussed in \cite{PaoloCF,Khoury1,MehrdadGrad}
\ba
\displaystyle\Psi=\Phi&=&b_ix^i \left[ 1-\dfrac{H}{a}\int dt'a(t') \right]\,,\nn\\ \label{4}
\displaystyle-\dfrac{\delta \rho}{\dot{\rho}}=\delta u&=&-b_ix^i\dfrac{1}{a}\int dt'a(t')\,,\\ \nn
\zeta= {\cal R}&=&-b_ix^i\,.
\ea

%%%%%%%%%%%%%%%%%%%%%%%%%%%%%%%%%%%%%%%%%%%%
\item \textbf{Gradient tensors (SCT)} Consider turning on $\sigma^0_{ijk}$ and the associated time dependent translation \eqref{ciit}
\ba
\displaystyle c_i&=&-\dfrac{1}{3}\sigma^0_{kki}\left[ \int \dfrac{dt'}{a(t')^3}\int^{t'} dt''a(t'') \right] \,.
\ea
Given the constraints summarized in table \ref{syms}, $\sigma^0_{ijk}$ has $ (6\times3)-3=15  $ independent components. We have already counted 3 of them, namely the SCT specified by a constant vector $b_i$ in \refeq{SCT}. The remaining 12 elements give rise to 12 gradient modes of the graviton
\ba\label{5}
\gamma_{ij}&=&-(\sigma^0_{ijl}+\sigma^0_{jil}+2\sigma^0_{lkk}\delta_{ij})x^l\,,\\ \nn
\displaystyle \Phi=\Psi &=&\dfrac{1}{3}\sigma^0_{kkl}x^l \Big(1-\dfrac{H}{a}\int dt' a(t')\Big)\,,\\ \nn
\displaystyle\dfrac{\delta \rho}{\dot{\rho}}= -\delta u &=&\dfrac{1}{3}\sigma^0_{kkl}x^l \left(\dfrac{1}{a}\int dt' a(t')\right)\,,\\ \nn
{\cal R}&=&-\dfrac{1}{3}\sigma^0_{kkl}x^l\,.
\ea 
%%%%%%%%%%%%%%%%%%%%%%%%%%%%%%%%%%%%
Again, the additional transversality requirement applies only for monochromatic modes and therefore it is relevant for the counting of independent soft theorems \cite{HinterbiCHLer:2013dpa}, but not for the counting of general adiabatic modes \cite{MehrdadGrad}. When one subtracts a SCT transformation with vector 
\be
b_i=-\dfrac{1}{3}\sigma^0_{kki}
\ee
from the above mode, one finds a pure gradient tensor mode \cite{PaoloCF}. 
\end{itemize}

%%%%%%%%%%%%%%%%%%%%%%%%%%%%%%%%%%%%%%%%%%%%%%%%%%%%%%%%%%%%
 
\subsection{New adiabatic modes}\label{ssec:}

We find a number of new adiabatic modes that, to best of our knowledge, have not been previously discussed in the literature. We report here these findings, starting with pure modes, namely modes with a well-defined tensor transformation under spatial rotations, and continuing with mixed modes.
%%%%%%%%%%
\begin{table}[H]
\small
\begin{center}
\begin{tabular}{|>{\columncolor[gray]{0.92}}c|c|}
\hlineB{3}
&\\ 
$\Phi=\Psi$& 
$
\begin{aligned}
&\left[-\dfrac{{\cal C}H}{a}+\dfrac{1}{3}\omega^0_{kk}(1-\dfrac{H}{a}\displaystyle\int dt'a(t'))+\dfrac{1}{3}\tilde{\omega}_{kk}(\int^t \dfrac{dt'}{a(t')^3}-\dfrac{H}{a}\int^t dt'a(t')\int^{t'}\dfrac{dt''}{a(t'')^3})\right]+\\
&\left[\dfrac{H}{a}({\cal C}_i-{\cal A}^0_i)+\dfrac{1}{3}\sigma^0_{kki}(1-\dfrac{H}{a}\displaystyle\int dt'a(t'))-\dfrac{1}{4}{\cal C}^0_{ikk}(\int^t \dfrac{dt'}{a(t')^3}-\dfrac{H}{a}\int^t dt'a(t')\int^{t'}\dfrac{dt''}{a(t'')^3})\right]x^i\\ 
&+\dfrac{1}{2}\left[-\dfrac{H}{a}({\cal B}^0_{ij}-\tilde{\omega}_{ij})-\dfrac{1}{4}{\cal D}^0_{ijkk}(\int^t \dfrac{dt'}{a(t')^3}-\dfrac{H}{a}\int^t dt'a(t')\int^{t'}\dfrac{dt''}{a(t'')^3})\right]x^ix^j\\ 
&+\dfrac{H}{6a}(\tilde{\sigma}_{ijk}-{\cal C}^0_{ijk})x^ix^jx^k+\dfrac{H}{24a}(\tilde{\mu}_{ijkl}-{\cal D}^0_{ijkl})x^ix^jx^kx^l
\end{aligned}
$
\\ 
&\\
\hlineB{3}
&\\
$\delta u$ & $ 
\begin{aligned}
&-\dfrac{{\cal C}}{a}-\dfrac{1}{3}\omega^0_{kk}\dfrac{1}{a}\int^t \dfrac{dt'}{a(t')}\\&+\dfrac{1}{3}\tilde{\omega}_{kk}\left(\dfrac{1}{\dot{H}a^3}-\dfrac{1}{a}\int^t dt' a(t')\int^{t'}\dfrac{dt''}{a(t'')^3}\right)\\ 
&+\left[\dfrac{{\cal C}_i-{\cal A}^0_i}{a}-\dfrac{1}{3}\sigma^0_{kki}\dfrac{1}{a}\int^t \dfrac{dt'}{a(t')}-\dfrac{1}{4}{\cal C}^0_{ikk}\left(\dfrac{1}{\dot{H}a^3}-\dfrac{1}{a}\int^t dt' a(t')\int^{t'}\dfrac{dt''}{a(t'')^3}\right)\right]x^i\\ 
&+\dfrac{1}{2}\left[\dfrac{\tilde{\omega}_{ij}-{\cal B}^0_{ij}}{a}-\dfrac{1}{4}{\cal D}^0_{ijkk}\left(\dfrac{1}{\dot{H}a^3}-\dfrac{1}{a}\int^t dt' a(t')\int^{t'}\dfrac{dt''}{a(t'')^3}\right)\right]x^ix^j\\ 
&+\dfrac{1}{6a}(\tilde{\sigma}_{ijk}-{\cal C}^0_{ijk})x^ix^jx^k+\dfrac{1}{24a}(\tilde{\mu}_{ijkl}-{\cal D}^0_{ijkl})x^ix^jx^kx^l
\end{aligned}
$
\\
&\\
\hlineB{3}
&\\
${\cal R}$ & $
\begin{aligned}
&-\dfrac{1}{3}\omega^0_{kk}+\dfrac{1}{3}\tilde{\omega}_{kk}(-\displaystyle\int \dfrac{dt'}{a(t')^3}+\dfrac{H}{\dot{H}a^3})-\dfrac{1}{3}\sigma^{0}_{kki}x^i+\dfrac{1}{4}{\cal C}^0_{ikk}(\displaystyle \int^t\dfrac{dt'}{a(t')^3}-\dfrac{H}{\dot{H}a^3})x^i\\ 
&+\dfrac{1}{8}{\cal D}^0_{ijkk}(\displaystyle \int^t\dfrac{dt'}{a(t')^3}-\dfrac{H}{\dot{H}a^3})x^ix^j
\end{aligned}
$\\ 
&\\
\hlineB{3}
&\\
$G_i$& 
$-\dfrac{1}{a^2}{\cal A}^0_i-\dfrac{1}{a^2}{\cal B}^0_{ij}x^j-\dfrac{1}{2a^2}{\cal C}^0_{ijk}x^jx^k-\dfrac{1}{6a^2}{\cal D}^0_{ijkl}x^jx^kx^l$\\ 
&\\
\hlineB{3}
&\\
$\delta u^V_i$ & $\dfrac{1}{4\dot{H}a^3}({\cal C}^0_{ikk}+{\cal D}^0_{ijkk}x^j)$\\
&\\
\hlineB{3}
&\\
$\gamma_{ij}$&
$
\begin{aligned}
&-2\omega^0_{(ij)}+\dfrac{2}{3}\omega^0_{kk}\delta_{ij}+\left(\dfrac{2}{3}\tilde{\omega}_{kk}\delta_{ij}-2\tilde{\omega}_{(ij)}\right)\displaystyle\int \dfrac{dt'}{a^3(t')}\\
&-\dfrac{1}{2}{\cal D}^0_{(ij)kk}\int^t \dfrac{dt'}{a(t')^3}\int^{t'}dt''a(t'')\int^{t''}\dfrac{dt'''}{a(t''')^3}-\left(\sigma^0_{ijl}+\sigma^0_{jil}+2\sigma^0_{lkk}\delta_{ij}\right)x^l\\
&-2\left(\dfrac{1}{4}{\cal C}^0_{lkk}\delta_{ij}+\tilde{\sigma}_{(ij)l}\right)\int \dfrac{dt'}{a^3(t')} x^l-\left(\dfrac{1}{4}{\cal D}^0_{lmkk}\delta_{ij}+\tilde{\mu}_{(ij)lm}\right)\int^t\dfrac{dt'}{a(t')^3}x^lx^m
\end{aligned}
$\\ 
&\\
\hlineB{3}
\end{tabular}
\caption{All of the adiabatic modes in an arbitrary FLRW universe, in which at least a zeroth order term or a gradient term appears.\label{diff} }
\end{center}
\end{table}
%%%%%%%%%%

\paragraph{Pure modes}

\begin{itemize}
\item \textbf{Time-dependent scalar mode} A time-dependent dilation with an appropriate temporal diff 
\ba
\ep_i=a^2 \dfrac{1}{3}\tilde{\omega}_{kk}\int^t\dfrac{dt'}{a(t')^3}x^i\quad , \quad \epsilon_0=\dfrac{1}{3a}\tilde{\omega}_{kk}\int^t a(t')dt'\int^{t'}\dfrac{dt''}{a(t'')^3}-\dfrac{1}{6a}\tilde{\omega}_{kk}\vec{x}^2\,,
\ea
generates a time dependent ${\cal O}(\vec{x}^0)$ mode in ${\cal R}$ with an ${\cal O}(\vec{x}^2)$ mode in the Newtonian potentials
\ba
\label{Newaddi}
{\cal R}&=&\dfrac{1}{3}\tilde{\omega}_{kk}\left(\dfrac{H}{\dot{H}a^3}-\int^t\dfrac{dt'}{a(t')^3}\right)\,,\\ \nn
\Phi&=&\Psi=\dfrac{1}{3}\tilde{\omega}_{kk}\left(\int^t \dfrac{dt'}{a(t')^3}-\dfrac{H}{a}\int^t dt'a(t')\int^{t'}\dfrac{dt''}{a(t'')^3}\right)
+\dfrac{H}{6a}\tilde{\omega}_{kk}\vec{x}^2\,.
\ea
This mode is present for a general fluid, but it disappears if the fluid under consideration is a (non-shift symmetric) scalar field. We discuss this in detail in Subsection \ref{rtdep}.
\item\textbf{Gradient of Weinberg's second scalar mode} A properly time-dependent spatial translation, accompanied by a temporal diff
\be
c_i={\cal C}_i\int \dfrac{dt'}{a^3(t')}\,,\quad \epsilon_0=-\dfrac{1}{a}{\cal C}_ix^i\,,
\ee
generates
\ba
\label{sgradd}
\Psi=\Phi&=&\dfrac{H}{a}{\cal C}_ix^i\,,\\ \nn
-\dfrac{\delta \rho}{\dot{\rho}}&=&\delta u=-\dfrac{{\cal C}_i}{a}x^i\then {\cal R}= 0 \,.
\ea
This mode can be interpreted in the same way as Weinberg's second adiabatic mode, namely as the difference between two pure gradient (SCT) modes.

%%%%%%%%%%%%%%%%%%%%%%%%%%%%%%%%%

\item \textbf{Vector mode} An appropriate time-dependent translation
\ba
c_i(t)&=&{\cal C}_i \int \dfrac{dt'}{a(t')^3}\,,\quad {\cal A}^0_i={\cal C}_i\,,
\ee
induces a pure vector mode
\be\label{vmode}
G_i=\dfrac{{\cal C}_i}{a^2}\,,\quad \delta u^{V}_{i}=0\,.
\ee
This new mode is discussed in detail in Subsection \ref{ssec:vector}.
%%%%%%%%%%%%%%%%%%%%%%%%%%%%%%%%%
\item \textbf{Gradient vector mode} Consider an antisymmetric ${\cal B}^0_{ij}={\cal B}^0_{[ij]}=\tilde \omega_{ij}$. Due to \eqref{Bij} this induces a time dependent rotation
\ba
\omega_{ij}&=&{\cal B}^0_{[ij]}\int \dfrac{dt'}{a(t')^3}\,,\quad\quad{\cal B}_{ij}(t)=0\,.
\ea
The resulting perturbation is the gradient of a pure vector mode (see Subsection \ref{ssec:vector})
\be
\label{grpv}
G_i=-\dfrac{1}{a^2}{\cal B}^0_{[ij]}x^j\,.
\ee
\end{itemize}

%%%%%%%%%%%%%%%%%%%%%%%%%%%%%%%%%

\paragraph{Mixed modes}

\begin{itemize}
\item\textbf{${\cal O}(\vec{x}^0)$ in tensor, ${\cal O}(\vec{x}^2)$ in scalars (or ${\cal O}(\vec{x}^0)$ in tensor, ${\cal O}(\vec{x})$ in vectors)} A symmetric and traceless $\tilde{\omega}_{ij}=\tilde{\omega}_{(ij)}$ corresponds to a time dependent anisotropic scaling, and it induces the following tensor-scalar adiabatic mode
\ba
\label{tendec}
\displaystyle \gamma_{ij}&=& -2\tilde{\omega}_{(ij)}\int \dfrac{dt'}{a(t')^3}\,,\\ \nn
\displaystyle\Psi=\Phi&=&\dfrac{H}{2a}\tilde{\omega}_{(ij)}x^ix^j\,,\\ \nn
\displaystyle\dfrac{\delta \rho}{\dot{\rho}}&=& -\delta u=-\dfrac{\tilde{\omega}_{(ij)}}{2a}x^ix^j\,.
\ea
The tensor part is recognized as the ``decaying'' tensor solution in standard expanding cosmologies. Remarkably, both time dependences of $  \gamma_{ij} $ are adiabatic
\be\label{teom}
\ddot \gamma_{ij}+3H\dot\gamma_{ij}=\mathcal{O}(q^{2}) \then \gamma_{ij}\sim\omega_{ij}^{0}-2\tilde \omega_{ij}\int\frac{dt}{a^{3}}\,.
\ee
According to the table \ref{diff}, by using
\be
\nn
{\cal B}^0_{ij}=-\tilde{\omega}_{(ij)}\,,
\ee
we can perform a temporal diff, which, when added to \eqref{tendec}, removes the scalar part and generates the following gradient term in vectors perturbations
\be
\label{mixv}
G_i=\dfrac{1}{a^2}\tilde{\omega}_{(ij)}x^j\,.
\ee
Notice that the decaying mode of tensor perturbations is adiabatic only when appropriately combined with a scalar (or vector) mode.  

%%%%%%%%%%%%%%%%%%%%%%%%%%%%%%%%%%%%%%%%%%%

\item\textbf{${\cal O}(\vec{x})$ in tensors, ${\cal O}(\vec{x}^3)$ in scalars (or ${\cal O}(\vec{x})$ in tensors, ${\cal O}(\vec{x}^2)$ in vectors)} A symmetric and traceless $\tilde{\sigma}_{ijk}$ matrix, yields twelve gradient modes of tensor type
\ba
\label{gt3s}
\sigma_{ijk}&=&\tilde{\sigma}_{ijk}\int \dfrac{dt'}{a(t')^3}\,,\\ \nn
\displaystyle \gamma_{ij}&=& -2\tilde{\sigma}_{ijk}x^k \Big(\int \dfrac{dt'}{a(t')^3}\Big)\,,\\ \nn
\Phi=\Psi&=&\dfrac{1}{6}\dfrac{H}{a}\tilde{\sigma}_{ijk}x^ix^jx^k\,,\\ \nn
\dfrac{\delta \rho}{\dot{\rho}}=-\delta u &=& -\dfrac{1}{6a}\tilde{\sigma}_{ijk}x^ix^jx^k\,.
\ea
We notice that the structure of $  \gamma_{ij} $ allows for a finite momentum extension that is monochromatic but also chromatic. Alternatively, we can also turn off scalars by choosing ${\cal C}^0_{ijk}=\tilde{\sigma}_{ijk}$, which in turn generates a number of quadratic vector modes
\be
G_i=\dfrac{1}{2a^2}\tilde{\sigma}_{ijk}x^jx^k\,.
\ee
%%%%%%%%%%%%%%%%%%%%%%%%%%%
\item\textbf{${\cal O}(\vec{x})$ in vectors, ${\cal O}(\vec{x}^2)$ in scalars (or ${\cal O}(\vec{x})$ in vectors, ${\cal O}(\vec{x}^0)$ in tensors)} The symmetric part of ${\cal B}^0_{ij}$ produces a mixed adiabatic mode as follows
\ba
\label{o1vo2s}
{\cal B}_{ij}&=&\dfrac{{\cal B}^0_{(ij)}}{a}\,,\quad
G_i=-\dfrac{{\cal B}^0_{(ij)}}{a^2}x^j\,,\\ \nn
\Phi &=&\Psi=H\delta u =-\dfrac{H}{2a}{\cal B}_{(ij)}x^ix^j\,.
\ea
It is evident that subtracting this mode from the first mixed adiabatic mode with a symmetric $\tilde{\omega}_{ij}=-{\cal B}^0_{(ij)}$ removes the scalar perturbations. This in turn produces an ${\cal O}(\vec{x}^0)$ tensor mode given by
\be
\gamma_{ij}=2{\cal B}^0_{(ij)}\int \dfrac{dt'}{a(t')^3}\,.
\ee
\item \textbf{${\cal O}(\vec{x}^0,\vec{x}^2)$ in vectors, ${\cal O}(\vec{x},\vec{x}^3)$ in scalars, ${\cal O}(\vec{x})$ in tensors} To construct this adiabatic mode, we choose a $\tilde{\sigma}_{ijk}$ matrix that is anti-symmetric in the first two indices. Because of the following symmetry property 
\be
\label{C0sym}
{\cal C}^0_{ijk}-\tilde{\sigma}_{ijk}=\text{totally sym.}\,,
\ee
one is obliged to choose an appropriate ${\cal C}^0_{ijk}$ matrix as well. 
If we choose the latter to be traceless over its last two indices, the following zeroth order perturbation in vorticity, and ${\cal O}(\vec{x}^2)$ perturbation in $G_i$ will be generated
\ba
\label{vmix}
\delta u^V_i=\dfrac{1}{4\dot{H}a^3}{\cal C}^0_{ikk}\quad \text{and}\quad 
G_i=-\dfrac{1}{2a^2}{\cal C}^0_{ijk}x^jx^k\,.
\ea
Because of the symmetries of ${\cal C}^0_{ijk}$ and $\tilde{\sigma}_{ijk}$ for this particular mode, scalar and tensor perturbations are inevitably present as well, and are given by
\ba
{\cal R}&=&\dfrac{1}{4}{\cal C}^0_{ikk}\left(\int^t\dfrac{dt'}{a(t')^3}-\dfrac{H}{\dot{H}a^3}\right)x^i\,,\\ \nn
\Phi&=&-\dfrac{1}{4}{\cal C}^0_{ikk}\left(\int^t\dfrac{dt'}{a(t')^3}-\dfrac{H}{a}\int dt'a(t')\int^{t'}\dfrac{dt''}{a(t'')^3}\right)x^i+\dfrac{H}{6a}(\tilde{\sigma}_{ijk}-{\cal C}^0_{ijk})x^ix^jx^k\,,\\ \nn
\gamma_{ij}&=&-\left(-\dfrac{1}{2}{\cal C}^0_{lkk}\delta_{ij}+2\tilde{\sigma}_{(ij)l}\right)\int^t \dfrac{dt'}{a(t')^3}x^l\,.
\ea
One may try to remove the tensor part by choosing a special  $\tilde{\sigma}_{ijk}$ with $\tilde{\sigma}_{(ij)l}=-\dfrac{1}{4}{\cal C}_{lkk}$. However, this is incompatible with the symmetries, i.e. using 
\eqref{C0sym} and ${\cal C}^0_{iik}=0$ one can see that such condition enforces ${\cal C}^0_{lkk}$ to vanish. 
\item \textbf{${\cal O}(\vec{x},\vec{x}^3)$ in vectors, ${\cal O}(\vec{x}^2,\vec{x}^4)$ in scalars, ${\cal O}(\vec{x}^2)$ in tensors}

A gradient mode in $\delta u_V^i$ can be constructed out of the ${\cal D}^0_{ijkl}$ matrix, with a non-vanishing trace over the last two indices. To be consistent with 
\be
{\cal D}^0_{ijkl}-\tilde{\mu}_{ijkl}=\text{totally sym.}\,,
\ee
an appropriate, although not unique, $\tilde{\mu}_{ijkl}$ matrix must be used. The final adiabatic mode consists of a gradient in vorticity together with an ${\cal O}(\vec{x}^3)$ in $G_i$
\ba
\label{vgmix}
\delta u^V_i=\dfrac{1}{4\dot{H}a^3}{\cal D}^0_{ijkk}x^j\quad \text{and}\quad G_i=-\dfrac{1}{6a^2}{\cal D}^0_{ijkl}x^jx^kx^l\,.
\ea
With such a choice of parameters, scalar and tensor perturbations become 
\ba
{\cal R}&=&\dfrac{1}{8}{\cal D}^0_{ijkk}\left(\int^t\dfrac{dt'}{a(t')^3}-\dfrac{H}{\dot{H}a^3}\right)x^ix^j\,,\\ \nn
\Phi&=&-\dfrac{1}{8}{\cal D}^0_{ijkk}\left(\int^t\dfrac{dt'}{a(t')^3}-\dfrac{H}{a}\int^t dt'a(t')\int^{t'}\dfrac{dt''}{a(t'')^3}\right)x^ix^j \\ \nn
&&+\dfrac{H}{24a}(\tilde{\mu}_{ijkl}-{\cal D}^0_{ijkl})x^ix^jx^kx^l\,,\\
\gamma_{ij}&=&-\dfrac{1}{2}{\cal D}^0_{(ij)kk}\int^t \dfrac{dt'}{a(t')^3}\int^{t'}dt''a(t'')\int^{t''}\dfrac{dt'''}{a(t''')^3}\\ \nn
&&-\left(\dfrac{1}{4}{\cal D}^0_{lmkk}\delta_{ij}+\tilde{\mu}_{(ij)lm}\right)\int^t\dfrac{dt'}{a(t')^3}x^lx^m\,.
\ea
%%%%%%%%%%%%%%%%%%%%%%%%%%%%%%%%%%
\end{itemize}
%%%%%%%%%%%%%%%%%%%%%%%%
\begin{table}[t]
\footnotesize
\begin{center}
\begin{tabular}{|c c c  |c|}
\rowcolor[gray]{0.9}&&&\\
\rowcolor[gray]{0.9}${\cal C}$&$\omega^0_{ij}$&
$\sigma^0_{ijk}$ &  Known\\
\rowcolor[gray]{0.9}&&&\\
&&&\\
$\checkmark$&&&$S^0$\\
&&&\\
&$\checkmark$&&$S^0$\\
&&&\\
&$\checkmark$&&$T^0$\\
&&&\\
&&$\checkmark$&$S^1$\\
&&&\\
&&$\checkmark$&$T^1$\\
&&&\\
\hline
\end{tabular}
%%%%%%%%%%%%%%%%%%%%%%
\begin{tabular}{|c c c c c  c c c|c|}
\hline
\rowcolor[gray]{0.9}&&&&&&&&\\
\rowcolor[gray]{0.9}${\cal C}_i$&${\cal A}^0_i$&${\cal B}^0_{ij}$&
$\tilde{\omega}_{ij}$& $\tilde{\sigma}_{ijk}$&${\cal C}^0_{ijk}$&$\tilde{\mu}_{ijkl}$&$\tilde{{\cal D}}^0_{ijkl}$&New \\
\rowcolor[gray]{0.9}&&&&&&&&\\
&&&&&&&&\\
&&&$\checkmark$&&&&&$S^{0,2}$\\
&&&&&&&&\\
$\checkmark$&&&&&&&&$S^1$\\
&&&&&&&&\\
&$\checkmark$&&&&&&&$V^0$\\
&&&&&&&\\
&&$\checkmark$&$\checkmark$&&&&&$V^1$\\
&&&&&&&&\\
&&($\checkmark$)&$\checkmark$&&&&&$T^0,S^2(V^1)$\\
&&&&&&&&\\
&&$\checkmark$&($\checkmark$)&&&&&$V^1,S^2(T^0)$\\
&&&&&&&\\
&&&&$\checkmark$&(\checkmark)&&&$T^1,S^3 (V^2)$\\
&&&&&&&&\\
&&&&$\checkmark$&$\checkmark$&&&$V^{0,2}, S^{1,3}, T^1$\\
&&&&&&&&\\
&&&&&&$\checkmark$&$\checkmark$&$V^{1,3}, S^{2,4}, T^{2}$\\
&&&&&&&&\\
\hline
\end{tabular}
\caption{In this table we have specified the Taylor coefficients used to build adiabatic modes. S,V and T stand for scalar, vector and tensor adiabatic modes, respectively. The $  n ,m$ integers specify that the mode is ${\cal O}(\vec{x}^{n})$ in one of the S, T or V metric and matter fields and  ${\cal O}(\vec{x}^{m})$ in the other. In parentheses alternative finite momentum interpretation of the same mode is shown.}
\end{center}
\end{table}

%%%%%%%%%%%%%%%%%%%%%%%%%%%%%%%%%%%%%%%%%%

%%%%%%%%%%%%%%%%%%%%%%%%%%%%%%%%%%%%%%%%
 
\section{Comoving gauge}\label{ssec:comoving}

In this section, we summarize the discussion of adiabatic modes in comoving gauge, defined by the condition
\bea
B=\delta u&=&0\,,\quad  C_i=0\,.
\eea
Notice that this gauge is only ``comoving'' with respect to the scalar component of the fluid velocity. The vector component $\delta u_i^V $, being small-gauge invariant, is arbitrary. At the cost of being pedantic, we introduce the variable $ \Rc $ as the curvature perturbation in comoving gauge\footnote{Of course $  \R $ is gauge-invariant and so, in value, $  \R=\Rc $. We introduce $  \Rc $ to avoid writing expressions for $  \R $ that are not manifestly gauge invariant.}
\be
\R|_{\text{comoving}}\equiv \Rc =\dfrac{A}{2}\,.
\ee
Again, we restrict ourselves to a single perfect fluid. In comoving gauge, the metric takes the form
\be
ds^2=-(1+2N_1)dt^2+2N_i dtdx^i+a^2 \left[ (1+2\Rc)\delta_{ij}+\gamma_{ij} \right]dx^idx^j\,,
\ee
where, to follow the common ADM notation, we renamed $  E=2 \No $ and $ G_{i}=\frac{1}{a}N_{i}  $.
Again $\gamma_{ij}$ is a transverse traceless tensor. We also need the scalar and vector parts of $N_i$, defined by
\be
\label{Ni}
N_i=\partial_i \psi+N_i^{V}\,,
\ee
where $  \partial_{i}N^{V}_{i}=0 $. By inverting the Laplacian, we can extract $\psi$
\ba
\label{psi}
\psi=\nabla^{-2}\partial_i N_i\,,
\ea
then $N_i^V$ can be read from \eqref{Ni}. The gauge transformation of perturbations are again given by \eqref{psi1}- \eqref{general}, and in comoving notation they read
\ba
\Rc &=& H\epsilon_0-\dfrac{1}{3a^2}\partial_k \epsilon_k\,,\quad N_{1} = \dot{\epsilon}_0\,,\quad \delta u_{i}=-\partial_i \epsilon_0\\ 
\dfrac{\delta \rho}{\dot{\rho}}&=& \epsilon_0\,,\quad N_i =-\partial_i \epsilon_0+2H \epsilon_i-\dot{\epsilon}_i\\
\gamma_{ij}&=&\dfrac{1}{a^2}\left(-2\e_{(i,j)}+\dfrac{2}{3}\partial_k\e_k\delta_{ij}\right)
\,.
\ea
Now we look for large diffeomorphisms that preserve the comoving gauge. We parallel\footnote{As pointed out in \cite{MehrdadGrad}, the final time dependence in (2.17) of \cite{HinterbiCHLer:2013dpa} is incorrect both because of a typo (a missing factor of $  a^{-2} $) and because it neglects contributions from the inverse Laplacian. In \cite{MehrdadGrad}, the authors could derive the correct time dependence by explicitly solving the equations of motion. Here we show that this is not necessary if one uses the $  ij $ Einstein equations, which are also satisfied trivially at $  q=0 $. This shows that it is still true that for adiabatic modes one does not need to solve any dynamical equation of motion.} the derivation of \cite{HinterbiCHLer:2013dpa} and derive adiabatic modes to all orders in $  x $ (in comoving gauge). This derivation clarifies the structure of the time dependence, which we will use in the next subsection to argue that adiabatic modes are classical. The Taylor expansion is postponed to the appendix \ref{app:comoving}.

The gauge parameters must satisfy \eqref{epi2}, \eqref{00c}, \eqref{save1}, \eqref{0ic} and \eqref{save2}, as discussed in appendix \ref{app:comoving}. 
According to \eqref{0ic} one can solve for $\epsilon_0$ in terms of $\e^i$, resulting in
\be
\e_0=\dfrac{1}{3\dot{H}}\partial_k \dot{\e}^k\,.
\ee
One should note that the gauge condition $\delta u=0 $ does not imply $\epsilon_0=0$ for two reasons. First, a spatially constant $\epsilon_0$ does not generate any perturbation in velocity. Second, for perturbations that do not fall off at infinity, a gradient term in $\delta u_i$, namely $-\partial_i \e_0$, can be absorbed into a pure vector mode (vorticity), because the separation between scalars and vectors is not unique at zero momentum. We can integrate \eqref{00c} to obtain the general form of $\psi$
\be
\label{defF}
\psi=\dfrac{F(\bx)}{a}-\dfrac{1}{3\dot{H}}\partial_k \dot{\e}^k+\dfrac{1}{3a}\int^t dt' a(t')\partial_k \e^k\,,
\ee
where we introduced $  F(\bx) $ as integration ``constant''.
Imposing \eqref{Nvcond} and \eqref{epi2}, we find
\ba
\label{epode}
N_i^V&\equiv&\dfrac{F_i(\bfx)}{a}=-a^2\dot{\e}^i-\dfrac{1}{3a}\int^t dt'a(t')\partial_i \partial_k \e^k-\dfrac{\partial_i F(\bx)}{a}\,,\\ \nn
0&=&\partial_i F_i\,,\\
\label{divepF}
0&=&\partial_k \dot{\e}^k+\dfrac{1}{a^3}\nabla^2 F\,.
\ea
From \eqref{epi2} and \eqref{divepF} we learn that $ \nabla^4 F=0 $. Plugging everything back in \eqref{epode} we find
\ba
\nn
\epsilon^i(t,x)&=&\bar{\epsilon}^i(\bx)-\left(\partial_i F+F_i\right)\int^t\dfrac{dt'}{a(t')^3}+\partial_i \nabla^2 F\int^t\dfrac{dt'}{3a(t')^3}\int^{t'}dt''a(t'')\int^{t''}\dfrac{dt'''}{a(t''')^3}\\
\label{t0}
&&-\partial_i \partial_k \bar{\e}^k\int^t \dfrac{dt'}{3a(t')^3}\int^{t'}dt'' a(t'')\,.
\ea
Taking a Laplacian from both sides reveals that the above solution satisfies \eqref{epi2}, iff i) $\bar{\e}^i(\bx)$ satisfies it and ii) the following relationship holds between $F$ and $F_i$
\be 
\nabla^2\left( F_i-\dfrac{4}{3}\partial_i F\right)=0\,.
\ee
Perturbations are then computed from the following general formulae
\ba
N_1&=&\dfrac{1}{3}\partial_t (\dfrac{1}{\dot{H}}\partial_k \dot{\e}^k)\,, \quad\quad{\cal R}_c=\dfrac{H}{3\dot{H}}\partial_k \dot{\e}^k-\dfrac{1}{3}\partial_k \e^k\,,\\ \nn
N_i&=&-\dfrac{1}{3\dot{H}}\partial_i \partial_k \dot{\e}^k-a^2 \partial_t \e^i\,,\quad\quad \gamma_{ij}=\dfrac{2}{3a^3}\partial_k \e_k \delta_{ij}-\dfrac{2}{a^2}\partial_{(i} \e_{j)}\,,\\\nn
 \delta u^V_i&=&-\dfrac{1}{3\dot{H}}\partial_i \partial_k \dot{\e}^k\,. 
\ea
Using 
\be
 \partial_{i}\nabla^{2}\bar \e^{i}=\nabla^{2}\nabla^{2}\epsilon^{i}=\partial_{i}F_{i}=\nabla^2 \nabla^{2}F=0 \,,
\ee
these perturbations simplify to\footnote{\label{ft2}The formula for $  \gamma_{ij} $ corrects (2.29) of \cite{HinterbiCHLer:2013dpa}, in agreement with \cite{MehrdadGrad}. As a check, one can verify that our $  \gamma_{ij} $ indeed solves its equations of motion, including the Laplacian term, as it should be
\be
\ddot \gamma_{ij}+3H\dot\gamma_{ij}-\frac{\nabla^{2}}{a^{2}}\gamma_{ij}=0\,.
\ee}
\be
{\cal R}_c&=&\left[ -\dfrac{H}{3\dot{H}a^3}+\dfrac{1}{3}\int^t \dfrac{dt'}{a(t')^3} \right]\nabla^2 F-\dfrac{1}{3}\partial_k \bar\e^k\,, \label{Newaddicg}\\ \nn
N_i&=&\frac{1}{a} \left(F_i(\vec{x})+\partial_i F(\vec{x})\right)+\dfrac{1}{3\dot{H}a^3}\partial_i \nabla^2 F-\dfrac{1}{3a}\partial_i \nabla^2 F\int^t dt'a(t')\int^{t'}\dfrac{dt''}{a(t'')^3}\\ \nn
&&+\dfrac{1}{3a}\partial_i \partial_k \bar{\e}^k\int^t dt'a(t')
\,, \label{R0}\\ 
\delta u^V_i&=&\dfrac{1}{3\dot{H}a^3}\partial_i \nabla^2 F\,, \quad \quad N_1=-\dfrac{1}{3}\nabla^2 F\partial_t(\dfrac{1}{\dot{H}a^3})\,,\\ \nn
 \gamma_{ij}&=&\dfrac{2}{3}\partial_k \bar\e^k \delta_{ij}-2\partial_{(i} \bar\e^{j)}-\dfrac{2}{3}\nabla^2 F\int^t \dfrac{dt'}{a(t')^3}\delta_{ij}+2\left(\partial_i \partial_j F+\partial_{(i}F_{j)}\right)\int^t \dfrac{dt'}{a(t')^3}\\ \nn
 &&-2\partial_i \partial_j \nabla^2 F \int^t \dfrac{dt'}{3a(t')^3}\int^{t'}dt''a(t'')\int^{t''}\dfrac{dt'''}{a(t''')^3}+2\partial_i \partial_j \partial_k \bar{\e}^k \int^t \dfrac{dt'}{3a(t')^3}\int^{t'}dt'' a(t'')\,.
\ee
In Appendix \ref{app:comoving} we have enumerated the leading adiabatic modes in the gradient expansion.

%%%%%%%%%%%%%%%%%%%%%%%%%%%%%%%%%%%%%%%%
 
\section{Properties of adiabatic modes}\label{ssec:prop} 
 
In this section, we discuss the new adiabatic modes we found, a number of general properties of all adiabatic modes and finally some subtleties in their physical interpretation.

%%%%%%%%%%%%%%%%%%%%%%%%%%%%%%%%%%%%%%%%

\subsection{New scalar adiabatic modes: Perfect fluid vs. generic scalar field}\label{rtdep}

In the last two sections, we discovered that for a generic single perfect fluid, the second time dependent mode of the curvature perturbation is adiabatic (\eqref{Newaddi} in Newtonian gauge and \eqref{Newaddicg} in comoving gauge). However, this is not the case for a generic\footnote{By ``generic'' we mean that it possesses no symmetries. As will be discussed elsewhere \cite{toappear}, if the scalar field possess a shift symmetry, this time-dependent scalar adiabatic mode survives and constrains, among other things, the correlators of Ultra Slow Roll inflation.} single scalar field! To see this, recall that
\ba\label{sol3}
\delta u=\dfrac{1}{3}\tilde{\omega}_{kk}\left(\dfrac{1}{\dot{H}a^3}-\dfrac{1}{a}\int^t dt' a(t')\int^{t'}\dfrac{dt''}{a(t'')^3}\right)+\dfrac{1}{6a}\tilde{\omega}_{kk}\bfx^2\,,
\ea
so while $  \partial_{i}\delta u=-\partial_{i} \e^{0}$, one finds $   \delta u\neq-\e_0$. For a generic fluid, this is allowed as long as we are dealing with large gauge transformations (while for small gauge transformations $\partial_{i}(\delta u+\e_0)=0$ would automatically imply $ \delta u=-\e_0  $). Things change dramatically for a generic scalar field because from the definition of the energy-momentum tensor one must impose the constraints
\be
\delta u^{V}_{i}=0\,,\quad \text{and}\quad\delta u_{i}=\partial_i\delta u+\delta u^{V}_{i}=-\dfrac{\partial_i\delta \phi}{\dot{\phi}}\,.
\ee
Imposing these constraints for perturbations that extend to finite momentum, one must have 
\be
\Delta \delta u=-\dfrac{\Delta\delta \phi}{\dot{\phi}}=-\e_0\,.
\ee

In fact, for a generic scalar field, any adiabatic mode for ${\cal R}$ is necessarily time independent. To prove this, recall that adiabatic modes in Newtonian gauge satisfy
\ba
\frac{\delta \rho}{\dot{\bar\rho}}&=&-\delta u\,,\quad\Phi=-\dot {\delta u}\,.\nonumber
\ea
Adiabatic modes in Newtonian gauge must also satisfy
\ba
\Phi&=&\Psi\,, \quad\dot \Phi=\dot H \delta u-H\Phi\,.\nonumber
\ea
From its gauge invariant definition, we then find
\ba
\dot \R&=&-\dot\Phi+\dot H \delta u+H\dot{\delta u}=-\dot H \delta u+H\Phi+\dot H \delta u+H\dot{\delta u}=0\,.\label{proof}
\ea
It is useful to understand how the Einstein Equations for a generic scalar field forbid the solution \eqref{sol3}. The details are presented in Appendix \ref{details}, while here we give an executive summary. In order to derive a dynamical equation of motion for ${\cal R}$ in comoving gauge, one needs a relationship between $\delta \rho$ and $\delta p$. For a generic adiabatic fluid such a relation in the $q\to 0$ limit is
\be
\label{adlim}
\dfrac{\delta \rho}{\delta p}=\dfrac{\dot{\rho}}{\dot{p}}\,.
\ee
But for a scalar field, $\delta \rho$ and $\delta p$ are not independent from metric perturbations and (in comoving gauge) \eqref{adlim} must be replaced by $\delta \rho=\delta p$. These two different constraints lead to different solutions for ${\cal R}(t)$. 
 
%%%%%%%%%%%%%%%%%%%%%%%%%%%%%%%%%%%%%%%%%
\subsection{New (mixed) tensor adiabatic modes}\label{ssec:}

In this paper, we have shown an interesting difference between scalar perturbations $  \R $ and tensor perturbations $  \gamma_{ij} $. For single generic scalar field, only one of the two solutions of the equations of motion for $  \R $ on large scales is adiabatic, namely $  \R= $ const. The second, time-dependent solution is adiabatic in general, but disappears if the fluid in question is a generic scalar field. This is in contrast with tensors, for which both solutions of the large-scale tensor equations of motion, namely
%\footnote{As noticed in footnote \ref{ft2}, the zero-momentum tensor modes \eqref{t0} solve the full equation of motion including the Laplacian term. On the other hand the zero-momentum $  \R $ in \eqref{R0} does not solve the full Mukhanov-Sasaki equation including the Laplacian. The reason is that the latter is obtained by integrating out the lapse $  N $ and shift $ N_{i}  $ from the Hamiltonian and momentum constraints. The solution, say (2.18) in \cite{HinterbiCHLer:2013dpa}, assumes that $  N $ and $  N_{i} $ vanish as spatial infinity, which is not the case for the solution in \eqref{R0}.}
\be\label{teom}
\ddot \gamma_{ij}+3H\dot\gamma_{ij}=\mathcal{O}(q^{2}) \then \gamma_{ij}\sim\text{const}+\int\frac{dt}{a^{3}}\,,
\ee
are adiabatic (up to $  \mathcal{O}(q^{2}) $ corrections). The first solution, which is constant up to order $  q^{2} $ corrections, is adiabatic by itself. Instead, the second solution is adiabatic only together with a scalar or vector perturbation. The fact that both leading order in $  q $ solutions can be made adiabatic is another way in which the properties of tensor modes are more robust than those of scalars, a fact recently emphasized in \cite{Bordin:2016ruc}. Tensor soft theorems so far have been derived for just the first adiabatic tensor mode (see e.g. \cite{Maldacena:2002vr,PaoloCF,HinterbiCHLer:2013dpa,Bordin:2016ruc}), generated by $  \bar \e^{i} $ in \eqref{t0}. The second solution is a decaying mode in standard expanding universes and so it usually subleading with respect to the first mode, but the opposite happens in contracting universes, such as the matter bounce \cite{Finelli:2001sr}. We discuss these soft theorems in Section \ref{ssec:soft}.

%This means that any quantum behavior of a pure $  \R $ mode cannot be capture via the Ward identities of the $  \R $ adiabatic modes. We speculate that therefore one should be able to derive all Ward identities of a pure $  \R $ adiabatic mode from a background wave method as in \cite{Maldacena:2002vr,PaoloCF}.

%%%%%%%%%%%%%%%%%%%%%%%%%%%%%%%%%%%%%%%%
 
\subsection{Vector adiabatic modes}\label{ssec:vector}

Since the existence of vector adiabatic modes is one of the new results of this work, we discuss it more in detail in this subsection. The nature of these modes is best understood by progressively enriching the dynamics of vector, from the case of a generic scalar field, where they are identically vanishing, to the most generic case of fully dynamical vectors. 

\paragraph{Scalar fields} Vector modes would be absent if, instead of a generic fluid, one considered a less general form of matter that does not admit vectorial deformation. An obvious example is a scalar field. It is instructive to see how our vector adiabatic modes disappear in this case. For a scalar field $  \delta u^{V}_{i}=0 $ identically. Then, the $ 0i  $ Einstein equation \refeq{0i-v} becomes trivial in $  q=0 $ zero limit since it reads in any gauge (see 5.1.51 of \cite{WB})
\be\label{veceq}
\nabla^2 \left( G^{V}_i -a\dot C^{V}_{i}\right)=0\quad\quad \text{(scalar field)}\,,
\ee
where the combination in parenthesis is gauge invariant. Then for scalar fields one needs to check that this equations is satisfied non-trivially at $  q\neq 0 $, namely $  G^{V}_{i}=a\dot C^{V}_{i} $. This means that the only gauge invariant vector perturbation, reducing to just $  G_{i} $ in the gauge we used in this paper, is identically zero and so no vector adiabatic modes are allowed. A scalar field was indeed assumed in \cite{HinterbiCHLer:2013dpa}, which explains why they did not find any vector adiabatic modes.

\paragraph{General perfect fluids} A general fluid admits vector perturbations $  \delta u^{V}_{i}\neq 0$. The $ 0i  $ Einstein equation \refeq{0i-v} now reduces to  
\be\label{vicina}
-4\dot{H}a \delta u^V_j=\nabla^2 \left( G^{V}_i -a\dot C^{V}_{i}\right)\,,%=0+\mathcal{O}\left( \frac{q^{2}}{a^{2}H ^{2}} \right)\,,
\ee 
As usual for adiabatic modes, we need to check only those equations that are trivially satisfied at $  q=0 $ because they are multiplied by some overall positive power of $  \vec{q} $. Since for pure adiabatic vector modes $  \delta u^{V}_{i}=0 $, one might be confused as to whether \eqref{vicina} is trivial at zero momentum or not. It is not trivial because the left-hand side does not vanish in general. Indeed, \eqref{vicina} is automatically satisfied both by the mixed vector modes, \eqref{vmix}, and by the pure adiabatic vector modes, \eqref{vmode}, up to $  q^{2} $ corrections. To understand this better, recall that for any (physical) adiabatic mode $  X $, one expects the following structure in the infrared:
\be\label{close}
X(\vec{q},t)=X^{\text{diff}}(t)+q^{2}X^{\text{phys}}(t)+\mathcal{O}(q^{4})\,
\ee
where $ X^{\text{diff}}  $ is the generated by some residual large diff and $ X^{\text{phys}}  $ depends on the dynamics of the system. As long as $  q $ is sufficiently small, the second term is a small correction to the first and the time dependence of $  X $ is well approximated by that of $  X^{\text{diff}} $. The case of pure vector modes for a perfect fluid obeys this structure but in a way that is doubly pathological. First, the $  \delta u^{V}_{i} $ generated by a diff, the equivalent of $  X^{\text{diff}} $ in \eqref{close}, vanishes and so $  \delta u^{V}_{i} $ starts directly at order $  q^{2} $. There is no limit in which the time dependence of $  \delta u^{V} $ is well described by a change of coordinates. Second, the other gauge invariant variable in the problem, namely the combination \eqref{veceq} of metric vector perturbations, does not receive any non-adiabatic correction. The equivalent of $  X^{\text{phys}} $ in \eqref{close} for $  G^{V}_i -a\dot C^{V}_{i} $ vanishes. The full linear solution of the $  ij $ trace-reversed Einstein equations \cite{WB}
\be
\partial_{k}\left( \partial_{t}+2H \right)\left( G^{V}_i -a\dot C^{V}_{i}\right)=0
\ee
coincides with the solution generated by the large diff, \eqref{vmode} and \eqref{grpv}. Both of these peculiarities of pure vector modes are accidents of perfect fluids and disappear for viscous fluids.

\paragraph{Viscous fluids} Consider a viscous fluid with energy-momentum tensor
\ba
T_{\mu\nu}&=&(\rho+p)u_\mu u_\nu+p g_{\mu\nu}\\ \nn
&& -2\eta \left(\nabla_{(\nu}u_{\mu)}+u^\kappa \nabla_{\kappa}u_{(\mu}u_{\nu)}\right)\\ \nn
&&-(\zeta-\dfrac{2}{3}\eta)\nabla_\kappa u^\kappa \left(g_{\mu\nu}+u_\mu u_\nu\right)\,,
\ea
where $  \zeta $ and $  \eta $ are the bulk and shear viscosity, respectively. In App \ref{app:viscous}, we show that the vector anisotropic stress produced by this system is 
\be\label{defpiv}
\pi_i^V =\dfrac{1}{a^2}\eta \left(aG_i-a^2\dot{C}_i-\delta u_i^V\right)\,,
\ee
which is gauge invariant as expected. The $  ij $ trace-reversed Einstein equations are now \cite{WB}
\be
\partial_{k}\left[ 16\pi G a \pi_{i}^{V}+\left( \partial_{t}+2H \right)\left( G^{V}_i -a\dot C^{V}_{i}\right) \right]=0\,,
\ee
while the second vector equation is unchanged, \eqref{veceq}. Using \eqref{veceq} and \eqref{defpiv} we see that the presence of some vector anisotropic stress induces a $  q^{2} $ correction to $  G^{V}_i -a\dot C^{V}_{i} $, whose time dependence depends on the physical properties of the system, such as viscosity, and cannot be derived from a change of coordinates, in accordance with the general expectation \eqref{close}. It would be interesting to go beyond fluids and study a system with a genuine vectorial dynamics, such as for example Electrodynamics. This might also be relevant for using our results in models of magnetogenesis. We leave this for future investigation.

%%%%%%%%%%%%%%%%%%%%%%%%%%%%%%%%%%%%%%%%
 
\subsection{Classical and quantum adiabatic modes}\label{sec:}

As we have just seen, there two cases in which adiabatic modes have fixed time dependence: pure vector adiabatic modes in general, $  N_{i}^{V} \propto a^{-1} $ and, in the specific but relevant case of a generic scalar field, scalar modes $  \R \sim $ const. In these cases adiabatic modes are classical in the following sense (also see \cite{Martin:2015qta,Martin:2017zxs}). Let us compute the only commutator that has a chance of being non-vanishing in the quantum theory of $  \R $ modes (a completely analogous argument applies to $  N_{i}^{V} $ mutatis mutandis), namely
\be
[\R(\vec{x},t),\dot \R(\vec{x},t)]\,.
\ee
Since we have been studying adiabatic modes at linear order, which are symmetries of the free theory\footnote{Adiabatic modes can be extended to symmetries of the full theory by including transformations that are linear in the fields \cite{Khoury1}. This is important to generate soft theorems to higher order in the soft moments, but we do not pursue it here.}, we can focus on free quantum fields
\bea
\R(\vec{x},t)&=&\int_{\vec{q}}\,\left[ e^{i\vec {q}\cdot \vec{x}}\,a_{\vec{q}}\,\R_{q}(t)+ e^{-i\vec {q}\cdot \vec{x}}\,a_{\vec{q}}^{\dagger}\,\R_{q}(t)^{\ast} \right]\\
&=& \int_{\vec{q}}\,e^{i\vec {q}\cdot \vec{x}} \R(\vec{q},t)\,.\label{otherf}
\eea
In terms of the absolute value and the argument $  \R_{q}(t)=|\R_{q}(t)|e^{i\theta_{q}(t)} $, the commutator is found to be
\be\label{short}
[\R(\vec{x},t),\dot \R(\vec{x},t)]=-2i\int_{\vec{q}}\,|\R_{q}(t)|^{2}\partial_{t}\theta_{q}(t)\,.
\ee
If the argument $  \theta_{q}(t) $ is time independent, the mode behave classically in the sense that all operators commute with each other\footnote{This is familiar from the heuristic discussion of classicalization of inflationary perturbations. In standard attractor slow-roll inflation, the commutator $   [\R,\dot\R]$ is proportional to the decaying mode because, when this is negligible, $  \R $ and $  \dot\R $ are proportional to the same quantum operator and therefore commute.}. This is indeed what happens for adiabatic modes with a single time dependence such as $  \R $, which is locally
\be
\R(\vec{x},t)=\sum_{n}^{\infty} a_{i_{1}i_{2}\dots i_{n}}x^{i_{1}}x^{i_{2}}\dots x^{i_{n}}\,.
\ee
By equating this Taylor expansion to the Taylor expansion around $  \vec{x}=0 $ (of course any other point will do since the difference is a time-independent phase) of \eqref{otherf}, one finds $  \partial_{t}|\R_{q}(t)|=\partial_{t}\theta_{q}(t)=0 $. For $  N_{i}^{V} $ the absolute value is instead time dependent, $  |N_{i}^{V}|\propto a^{-1} $, but its argument is still time independent, so the same conclusion holds.

Things are different for adiabatic perturbations with two different time dependences, such as tensor modes $  \gamma_{ij} $ and scalar modes for generic fluids. This situation has not been considered previously in the literature because, to the best of our knowledge, we are the first to point out the adiabatic nature of the decaying tensor modes. Around an arbitrary point which we choose to be the origin, tensor adiabatic perturbations have the general structure
\be\label{2tdep}
\gamma(\vec{x},t)=\sum_{n}^{\infty} a_{i_{1}i_{2}\dots i_{n}}x^{i_{1}}x^{i_{2}}\dots x^{i_{n}}+\int \frac{dt}{a^{3}} \,\sum_{n}^{\infty} b_{i_{1}i_{2}\dots i_{n}}x^{i_{1}}x^{i_{2}}\dots x^{i_{n}}\,,
\ee
where we omitted the indices of $  \gamma $ because they are irrelevant for this discussion and the $  a $'s and $  b $'s are some constants (not to be confused with the $  b $ we used for SCT). Since tensor modes obey a second order differential equation in time, we conclude that, up to $  \mathcal{O}(q^{2}) $ corrections, a generic tensor mode, \textit{including its decaying mode} can be approximated by a gauge transformation. We can write
\be
\gamma(\vec{x},t)&=&\int_{\vec{q}}\,e^{i\vec {q}\cdot \vec{x}} \left[ \gamma^{a}(\vec{q},t)+\gamma^{b}(\vec{q},t) \right]\,,\\
\gamma^{a,b}(\vec{q},t)&=&\gamma^{a,b}_{q}(t)a_{\vec{q}}+\gamma^{a,b}_{q}(t)a_{-\vec{q}}^{\dagger}\,,\\
\gamma^{a,b}_{q}(t)&=&|\gamma^{a,b}_{q}(t)|\exp\left[  i\theta_{q}^{a,b}(t) \right]\,,
\ee
where $  \gamma^{a,b}(\vec{q},t) $ are the Fourier transform of each of the two time dependences in \refeq{2tdep}. From the previous discussion of $  \R $ and $  N^{V}_{i} $, we know that $  \partial_{t}\theta_{q}^{a,b}(t)=0 $. We can then use \eqref{short} to write
\be
[\gamma(\vec{x},t),\dot\gamma(\vec{x},t)]=-2i\int_{\vec{q}}\,|\gamma^{a}_{q}(t)+\gamma^{b}_{q}(t)|^{2}\partial_{t}\left[ \text{Arg} \left(  \gamma^{a}_{q}(t)+\gamma^{b}_{q}(t)\right)\right]\,.
\ee
The time derivative can be computed to be
\be
\partial_{t}\left[ \text{Arg} \left(  \gamma^{b}+\gamma^{b}\right)\right]=\frac{\sin(\theta^{a}-\theta^{b}) \left[  |\dot\gamma^{a} \gamma^{b}|-|\gamma^{a}\dot \gamma^{b}|\right]}{|\gamma^{a}|^{2}+2\cos\left(  \theta^{a}-\theta^{b}\right)|\gamma^{a}\gamma^{b}|+|\gamma^{b}|^{2}}\,,
\ee
where we suppressed the time and $  q $ dependence. This quantity is in general non-vanishing and hence so is the commutator. We conclude that adiabatic modes with two time dependences, such as for example general adiabatic tensor modes, are quantum in the sense specified above. This is in contrast to scalar adiabatic modes of curvature perturbations $  \R $ in the presence of a generic scalar field.

Physical modes such as scalar, vector and tensors arising from quantizing perturbations in FLRW spacetime are approximated by adiabatic modes only locally, or equivalently in Fourier space only up to $  q^{2} $ corrections. They differ from their adiabatic approximation by some non-adiabatic corrections whose effect is not a local diffeomorphism. The time dependence of the non-adiabatic part differs in general from that of the adiabatic part, and therefore contribute to a non-vanishing commutator.

%%%%%%%%%%%%%%%%%%%%%%%%%%%%%%%%%%%%%%%%
 
\subsection{The third man: on the nature of Weinberg's second adiabatic mode}\label{thirdman}

It is worth understanding how Weinberg's second adiabatic mode (WAM II) continues to  its finite momentum counterpart. For WAM II  ${\cal R}_c$ vanishes, but a naive expectation suggests that if we keep the momentum finite and Taylor expand in $q$, then ${\cal R}_c$ receives its first correction at order $q^2$. To illustrate this point, using \eqref{prho} we derive a second order equation of motion for $\Phi$ in the Newtonian gauge: 
\be
\ddot{\Phi}+\left(H-\dfrac{\ddot{H}}{\dot{H}}\right)\dot{\Phi}+\left(2\dot{H}-\dfrac{H\ddot{H}}{\dot{H}}\right)\Phi-\left(1+\dfrac{\ddot{H}}{3H\dot{H}}\right)\dfrac{q^2}{a^2}\Phi=0\,.
\ee
This equation can be solved order by order in $q^2$. It is easy to explicitly check that:
\be
\Phi_q=\Phi_0+q^2\Phi_1+...=\dfrac{H}{a}-\dfrac{1}{3}q^2\left(\int^t \dfrac{dt'}{a(t')^3}-\dfrac{H}{a}\int^t dt'a(t')\int^{t'}\dfrac{dt''}{a(t'')^3}\right)+... \,,
\ee
where we neglected the second solution at order $  q^{0} $, which is not relevant for our discussion. The curvature perturbations is
\be
\label{q2R}
{\cal R}=-\dfrac{q^2}{3}\left(\dfrac{H}{\dot{H}a^3}-\int^t\dfrac{dt'}{a(t')^3}\right)+...\,,
\ee
which confirms our expectation that the curvature perturbations for this mode start at order $q^2$. Indeed, we can recover our new adiabatic mode for curvature perturbation by expanding the finite momentum version of WAM II in powers of $\vec{x}$. Since we will average over the direction of the mode $\vec{q}$ later on, we drop the $\sin(\vec{q}\cdot\vec{x})$ contribution here as it vanishes after averaging over $\hat{q}$. So we consider the following cosine mode
\ba
\label{finiteWAMII}
\Phi(t,\vec{x})&=&\left[\dfrac{H}{a}-\dfrac{1}{3}q^2\left(\int^t \dfrac{dt'}{a(t')^3}-\dfrac{H}{a}\int^t dt'a(t')\int^{t'}\dfrac{dt''}{a(t'')^3}\right)+... \right]\cos (\vec{q}\cdot\vec{x})\\ \nn
&=&\dfrac{H}{a}+q^2\left[-\dfrac{H}{2a}\hat{q}_i\hat{q}_jx^ix^j-\dfrac{1}{3}\left(\int^t \dfrac{dt'}{a(t')^3}-\dfrac{H}{a}\int^t dt'a(t')\int^{t'}\dfrac{dt''}{a(t'')^3}\right)\right]+...\,,\\ \nn
{\cal R}(t,\vec{x})&=& \left[-\dfrac{q^2}{3}\left(\dfrac{H}{\dot{H}a^3}-\int^t\dfrac{dt'}{a(t')^3}\right)+...\right]\cos(\vec{q}\cdot\vec{x})\\ \nn
&=&-\dfrac{q^2}{3}\left(\dfrac{H}{\dot{H}a^3}-\int^t\dfrac{dt'}{a(t')^3}\right)+\dfrac{q^4}{6}\left(\dfrac{H}{\dot{H}a^3}-\int^t\dfrac{dt'}{a(t')^3}\right)\hat{q}_i\hat{q}_jx^ix^j+... \,.
\ea
The above expansion performed in terms of $q$, must solve the Einstein equations, order by order. After averaging over $\hat{q}_i$, one can clearly see that the $q^2$ and $\vec{x}$ dependent part is indeed the first pure adiabatic mode that we discovered in \eqref{Newaddi}. In other words, WAM II and the new scalar adiabatic mode are the $q^0$ and $q^2$ behavior of the same physical isotropic mode, respectively. 

Another important by-product of the previous discussion is that the second time-dependent mode of curvature perturbations is not a pure adiabatic mode, unless averaged over the direction of the momentum. A monochromatic mode, on the other hand, would be a mixed adiabatic mode. A direct comparison between \eqref{finiteWAMII} and Table \ref{diff} reveals that a fixed $\vec{q}$ mode in the limit $q\to 0$ is adiabatic iff it is mixed with the following tensor mode,
\be
\tilde{\omega}_{ij}=-\hat{q}_i\hat{q}_j\Rightarrow \gamma_{ij}=\left(-\dfrac{2}{3}\tilde{\omega}_{kk}\delta_{ij}+2\hat{q}_i\hat{q}_j\right)\int^t \dfrac{dt'}{a(t')^3}\,. 
\ee
This tensorial perturbation can not be monochromatic because for the above $\gamma_{ij}$ matrix, the equation $\hat{p}_i \gamma_{ij}=0$ does not possess any solution.

What happens for a single, generic scalar field? In this case, as we have proven around \eqref{proof}, any adiabatic mode of ${\cal R}$ must be constant in time. Therefore, the time dependence in front of $q^2$ in \eqref{q2R} can not be fixed without solving the full dynamical equations of motion. That is the reason for the violation of Maldacena's consistency conditions in ultra-slow roll models \cite{Namjoo1,Namjoo2,Namjoo3,Mooij:2015yka,AkhshikSadra}. In other words, in situations in which this mode dominates over Weinberg's first adiabatic mode, there is no coordinate transformation locally resembling the long mode\footnote{The relevant exceptions of shift-symmetric scalar fields will be discussed elsewhere \cite{toappear}}. Moreover, we notice that WAM II is not associated with any Noether current because, after integrating out the shift $N_i$, the symmetry associated with WAM II disappears from the ${\cal R}_c$ Lagrangian. 
%For instance, in comoving gauge WAM II is equivalent to a gradient mode, as already pointed out in \ref{foot}. The diffeomorphism that generates it is
%\be
%\epsilon^i=\bar{\cal C}_i\int^t \dfrac{dt'}{a^3(t')}\Rightarrow \psi=-\dfrac{1}{a}\bar{\cal C}_i x^i\,.
%\ee

%%%%%%%%%%%%%%%%%%%%%%%%%%%%%%%%%%%%%%%%
 
\section{Soft theorems}\label{ssec:soft}

One of the most interesting consequences of cosmological adiabatic modes are the soft theorems derived as Ward identities of the associated symmetries. Here we provide a brief preliminary discussion the soft theorems implied by the new vector and tensor adiabatic modes we have found in this work. The soft theorem of the new time-dependent scalar mode of $  \R $ will be discussed elsewhere \cite{toappear}. We follow the background wave method of \cite{Maldacena:2002vr,PaoloCF}, but similar results can be derived from Ward identities \cite{Assassi:2012et,Assassi:2012zq,HinterbiCHLer:2013dpa,Berezhiani:2014tda}, 1PI methods \cite{Goldberger:2013rsa}, the wavefunction of the universe \cite{Pimentel:2013gza,Ghosh:2014kba,Kundu:2014gxa,Kundu:2015xta} or Slavnov-Taylor identities \cite{Berezhiani:2013ewa,Collins:2014fwa}.

%%%%%%%%%%%%%%%%%%%%%%%%%%%%%%%%%%%%%%%%

\subsection{Vector soft theorems}\label{ssec:}

Since we discovered adiabatic vector modes, we can derive the corresponding soft theorems (a.k.a. consistency conditions). Here we derive only the soft theorems corresponding to the pure vector adiabatic mode, for which $\delta u^V_i=0$, postponing the mixed case to future work. These soft theorems are expected to apply only in the absence of non-adiabatic superHubble vector perturbations, as it is the case for scalars. Hence we consider a universe filled with a single fluid, accompanied by other kinds of matter that do not admit vector perturbations. To be concrete, we present our results in Newtonian gauge. The local effect of a long vector mode $G_i$, up to leading order in $q_L$, is a time-dependent translation, 
\ba
\nn
\epsilon_0&=&0\,,\\ 
\epsilon^i&=&-a^2G_i(\vec{q})\int \dfrac{dt'}{a(t')^3}\,.
\ea
Let us look at the correlation between a soft vector mode and the product of any other ``hard''  perturbations. Using the background wave method we find
\ba
\nn
\lim_{q\to 0}\langle G_i(\vec{q},t){\cal O}(\vec{k}_1,... \vec{k}_n,t)\rangle
&=&\int d^3x_1.. d^3x_n e^{-i\vec{k}_1.\vec{x}_1}...e^{-i\vec{k}_n.\vec{x}_n}\left\langle G_i(\vec{q},t)\langle {\cal O}(\vec{x}_1,..,\vec{x}_n;t)\rangle_{G_i(\vec{q}')}\right\rangle \\ \nn
&=&\int d^3x_1.. d^3x_n e^{-i\vec{k}_1.\vec{x}_1}...e^{-i\vec{k}_n.\vec{x}_n}\\ \nn
&&\times\left\langle G_i(\vec{q},t) \langle{\cal O}\left(\vec{x}_1+\vec{G}(\vec{q},t),... ,\vec{x}_n+\vec{G}(\vec{q},t)\right)\rangle \right\rangle\\ 
&=&P_{\vec{G}}(q,t)\vec{e}(\hat{q}).(\vec{k}_1+...+\vec{k}_n) \langle{\cal O}(\vec{k}_1,... \vec{k}_n,t)\rangle={\cal O}(q^2)\,.
\ea
where $\vec{e}(\hat{q})$ is the polarization vector associated with the soft mode. This result was expected since a time-dependent translation should not affect equal-time correlations because of statistical homogeneity. It would be interesting to compute the effect for unequal time correlators, especially in view of applications to large scale structures, along the lines of \cite{Creminelli:2013mca,Peloso:2013zw,Kehagias:2013yd}.

%%%%%%%%%%%%%%%%%%%%%%%%%%%%%%%%%%%%%%%%
\subsection{Tensor soft theorems}\label{ssec:}

In an expanding universe, the time-dependent mode of the graviton decays with time. However, in contracting universes this mode grows and ultimately dominates over the constant one. On superHubble scales, we can write the evolution of $\gamma_{ij}$ as
\be
h_{ij}(q,t)=c_1+c_2 \int^t \frac{dt'}{a(t')^{3}}\,.
\ee
Since $  \dot h \propto a^{-3} $, in expanding universes this mode eventually becomes constant, i.e. the $  c_{1} $ solution dominates in the future. The opposite happens in contracting universes, where $  c_{2} $ dominates.
%where $a_{\ast}$ and $H_{\ast}$ are the scale factor and Hubble expansion rate at horizon crossing, i.e. $q=a_{\ast}H_{\ast}$. 
%Those backgrounds in which every mode exit the horizon and before, it has been in its Minkwski vaccum, it seems reasonable to assume $\dfrac{c_1}{c_2}={\cal O}(1)$. Which is to say that the positive frequency has same order amount of contribution from both modes. If we ask the time dependent peice to vanquish the constant peice we then must have: 
%\be
%\label{req}
%\int^{t_e}_{t_{hc}}dt'H_{hc}\dfrac{a_{hc}^3}{a(t')^3} \gg 1\,,
%\ee
%Where $t_e$ stands for the end of the contracting phase, i.e. where we calculate all of the correlation functions. As an illustration, 
%
%In an expanding universe $a(t)$ is monotonic so we might write 
%\be
%\int^{t_e}_{t_{hc}}dt'\dfrac{a_{hc}^3 H_{hc}}{a(t')^3}=\int^{a_e}_{a_{hc}}\dfrac{H_{hc}da}{a^3(aH)}\,.
%\ee
%In a universe where modes exit the Horizon, $\frac{1}{aH}$ is decreasing (i.e. $\frac{1}{aH}<\frac{1}{a_iH_i}$)as a result above integral is less than unity. 
For example, in the matter bounce considered in \cite{Battefeld:2004cd}, one finds
\be
a(t)\propto t^{2/3}\Rightarrow h_{ij}(t\rightarrow 0_{-})\propto \frac{c_{2}}{t}\,,
\ee
where time runs from $  -\infty  $ to $  0 $. In these scenarios, there exist soft theorems associated with the second, time-dependent adiabatic tensor mode. In comoving gauge and to leading order in the gradient expansion, an adiabatic, time-dependent tensor locally resembles a time-dependent anisotropic scaling
\be 
\epsilon^i=-\dfrac{1}{2}\gamma_{ij}(t)x^j\,.
\ee
This coordinate transformation is the same as the one for the constant tensor adiabatic mode. Since for this mode $\epsilon_0=0$, the time-dependence above does not change the soft theorems. As a result, all the soft theorems involving a single long graviton in a contracting universe are the same as in the ordinary expanding universes \cite{Maldacena:2002vr}\footnote{The same statement also applies to ${\cal O}(q)$ soft theorems involving a soft graviton.}.

%%%%%%%%%%%%%%%%%%%%%%%%%%%%%%%%%%%%%%%%

\section{Conclusion}\label{concl}

In this work, we provided a systematic derivation of cosmological adiabatic modes in flat FRLW spacetime, extending previous results \cite{WAD,Khoury1,HinterbiCHLer:2013dpa}. By allowing for vector perturbations in the fluid, we discovered new vector adiabatic modes and many mixed modes, including the time-dependent tensor mode. These are all decaying modes in expanding universes, while they grow and dominate in contracting backgrounds. In addition, we found that the time-dependent mode of curvature perturbations ${\cal R}$ is adiabatic for a general perfect fluid. Remarkably, this mode loses its adiabatic character if the fluid is a generic scalar field. We also discussed the implications of the new tensor and vector adiabatic modes for cosmological soft theorems. The time-dependent tensor mode dominates in contracting universes, but the leading local effect of the long tensor mode is exactly the same as the would-be-freezing mode, namely an anisotropic scaling proportional to the graviton amplitude. Consequently, the known soft theorems that hold for soft gravitons in standard cosmology should hold for contracting universes as well. 
 
Our work can be extended in many directions:
 \begin{itemize}
 \item One should extend our systematic study to spatially curved universes. At any non-vanishing order in the spatial curvature $  K $, scalar modes are not adiabatic anymore and therefore the related soft-theorems should be violated (see also \cite{Sugimura:2013cra}). On the one hand, observations tells us that ${\Omega}_k<10^{-3}$ \cite{Ade:2015xua}, and so this is probably subleading. On the other, the squeezed limit of three point function can be tested to extremely high precision with futuristic measurement of CMB spectral distortion \cite{Pajer:2012vz} and so this effect might be measurable after all.
 \item One should extend our preliminary analysis of the soft theorems corresponding to the new adiabatic modes we discussed here. This can be relevant to characterize alternatives to inflation that postulate a primordial phase of contraction, such as the matter bounce \cite{Battefeld:2004cd,Cai:2009fn,Li:2016xjb}, the ekpyrotic universe \cite{Khoury:2001wf} or the galilean genesis \cite{Creminelli:2010ba}. 
 \item It would be interesting to see if the new vector adiabatic modes can play a role for primordial magnetogensis, either from inflation or its alternatives. For a discussion of vector modes in contracting universes see \cite{VecP}. 
 \item In this paper we assumed that the fluid does not support any anisotropic stresses on large scales. For mundane fluids such as particle Dark Matter, baryonic matter or decoupled neutrinos, this is a very good approximation. However, in principle one can consider peculiar matter that enjoy substantial amount of anisotropies on super Hubble scales, such as for example solid inflation \cite{Endlich:2012pz}. This requires including in our derivation a non-trivial $\pi_{ij}$ and its transformations under large diffeomorphisms. This problem might also be related to magnetogenesis scenarios during inflation, as they need anisotropic stresses on large scales \cite{MukhB}.
 \item The separate universe approximation is a closely related approach for studying cosmological perturbations on large scales (for an incomplete list of references see \cite{LythMalik1, WandsMalik, LythSasaki, Talebian}). It would be interesting to compare our results with those derived using that method.
 \item Given that all scalar adiabatic modes satisfy $\nabla^2 \R=0$, it would be interesting to investigate the relation to the cosmological zero modes studied in \cite{Afshordi:2017use}.
 \end{itemize}
 
 %%%%%%%%%%%%%%%%%%%%%%%%%
 
 \section*{Acknowledgments} 
We would like to thank Bernardo Finelli, Sander Mooij, Mehrdad Mirbabayi, Riccardo Rattazzi, John Stout, Dong-Gang Wang, Yvett Welling and Drian van der Woude for useful comments and discussions. We are particularly thankful to Guus Avis, Garrett Goon and Luca Santoni for comments on the draft. E.P. and S.J. are supported by the Delta-ITP consortium, a program of the Netherlands organization for scientific research (NWO) that is funded by the Dutch Ministry of Education, Culture and Science (OCW). S.J. would like to thank Hassan Firouzjahi and Ali Akbar Abolhasani for useful discussions and continuous encouragements. S.J. also thanks Utrecht University for the warm hospitality during the completion of this project. The work of S.J. is also supported by the Iranian National Elites Foundation(BMN).   
 
 %%%%%%%%%%%%%%%%%%%%%%%%%%%%%%%%%%
 
 \appendix

%%%%%%%%%%%%%%%%%%%%%%%%%%%%%%%%%%%%%%%%
 
\section{Gradient expansion}\label{app:det}

In this appendix we collect the calculation of the constraints imposed on the Taylor coefficients of gauge transformations in Newtonian gauge. The spatial gauge parameter $\epsilon_i$ can be expanded to cubic order in $\vec{x}$
\be
\label{ei}
\e^{i}=\dfrac{1}{a^2}\ep_i\simeq c_i(t)+\omega_{il}(t)x^l+\dfrac{1}{2!}\sigma_{ikl}(t)x^kx^l+\dfrac{1}{3!}\mu_{ijkl}x^jx^kx^l\,,
\ee
where, by \eqref{epi2}, $\sigma_{ijk}$ satisfies
\ba
\label{sig}
\sigma_{ijk}=\sigma_{ikj}\quad \text{and}\quad \sigma_{ijj}=-\dfrac{1}{3}\sigma_{jji}\,.
\ea
Analogously, for $\mu_{ijkl}$ we have
\be
\label{mu}
\mu_{ijkl}=\text{symmetric in jkl}\quad \text{and}\quad \mu_{ilkk}=-\dfrac{1}{3}\mu_{kkil}\,.
\ee

To satisfy \eqref{ep0} we need to keep $\epsilon_0$ up to quartic order in $  \vec{x} $
\be\label{e0}
\epsilon_0\simeq\epsilon(t)+{\cal A}_i(t) x^i+\dfrac{1}{2!}{\cal B}_{ij}(t) x^ix^j+\dfrac{1}{3!}{\cal C}_{ijk}(t) x^ix^jx^k+\dfrac{1}{4!}{\cal D}_{ijkl}x^ix^jx^kx^l\,,
\ee
where ${\cal B}_{ij}$, ${\cal C}_{ijk}$ and ${\cal D}_{ijkl}$ are totally symmetric and satisfy 
\ba
{\cal B}_{kk} &=& -a^2\dot{\omega}_{kk}\, ,\\ \nn
{\cal C}_{ikk} &=&-a^2 \dot{\sigma}_{kki}\,,\\ \nn
{\cal D}_{ijkk}&=&-a^2 \dot{\mu}_{kkij}\,,
\ea
Imposing \eqref{off-v} %and enforcing $G_i$ to be at most linear\footnote{This is to avoid additional diff's that are higher order in $  \vec{x} $.} in $\bfx$ 
we find 
\ba 
\label{Ai}
{\cal A}_i+a^2\dot{c}_i &=& \dfrac{{\cal A}^0_i}{a}\,,\\ 
\label{Bij}
{\cal B}_{ij}+a^2 \dot{\omega}_{ij}&=&\dfrac{{\cal B}^0_{ij}}{a}\,,\\
\label{Cijk}
{\cal C}_{ijk}+a^2 \dot{\sigma}_{ijk} &=& \dfrac{{\cal C}^0_{ijk}}{a}\,,\\ 
\label{Dijkl}
{\cal D}_{ijkl}+a^2\dot{\mu}_{ijkl}&=&\dfrac{{\cal D}^0_{ijkl}}{a}\,,
\ea
where $ {\cal A}^0_i  $, $ {\cal B}^0_{ij}  $, ${\cal C}^0_{ijk}$ and ${\cal D}^0_{ijkl}$ are time-independent integration constants. While the ${\cal A}^0_i$ is arbitrary, the others are subject to the constraint $\partial_i G_i=0$ and therefore
\be
{\cal B}^0_{ii}={\cal C}^0_{iik}={\cal D}^0_{iikl}=0 \,.
\ee
Since ${\cal B}_{ij}$ and ${\cal C}_{ijk}$ must be symmetric, \eqref{Bij} and \eqref{Cijk} further imply
\be
\label{totsymm}
{\cal B}^0_{[ij]}=a^3\dot{\omega}_{[ij]}\,,\quad{\cal C}^0_{[ij]k}=a^3\dot{\sigma}_{[ij]k}\quad \text{and} \quad{\cal D}^0_{[ij]kl}=a^3\dot{\mu}_{[ij]kl}\,.
\ee

From \eqref{0i-s} one can derive $\delta u$:
\be
\delta u=-\epsilon_0+\dfrac{1}{3\dot{H}}\partial_k \dot{\epsilon}^k\,.
\ee
As a result, from \eqref{psi} we learn that
\be
\label{VV}
\delta u^V_i=-\dfrac{1}{3\dot{H}}\partial_i \partial_k \dot{\epsilon}^k\,.
\ee

Finally, \eqref{off-s} fixes the time dependence of the $  \e^{i} $ Taylor coefficients in \eqref{ei} to
\bea
\label{siggt}
\sigma_{ijk}(t)&=&\tilde{\sigma}_{ijk}{\displaystyle \int \dfrac{dt'}{a(t')^3}}+\sigma^0_{ijk}\quad , \quad \tilde{\sigma}_{kki}=-3\tilde{\sigma}_{ikk}\,,\\
\label{mut}
\mu_{ijkl}&=&\tilde{\mu}_{ijkl}{\displaystyle \int \dfrac{dt'}{a(t')^3}}+{\mu}^0_{ijkl}\quad , \quad \tilde{\mu}_{iikl}=-3\tilde{\mu}_{klii}\,,\\ \nn
\omega_{ij}(t)&=&\tilde{\omega}_{ij}{\displaystyle\int \dfrac{dt'}{a(t')^3}+\omega^0_{ij}}-\dfrac{1}{3}\mu^0_{kkij}\int^t \dfrac{dt'}{a(t')^3}\int dt''a(t'')\\
\label{omega}
&&+\dfrac{1}{4}{\cal D}^0_{ijkk}\int^t \dfrac{dt'}{a(t')^3}\int^{t'}dt''a(t'')\int^{t''}\dfrac{dt'''}{a(t''')^3}\,,\\ 
\label{eppt}
\ep(t)&=&\dfrac{1}{3a}\tilde{\omega}_{ii}\int^t dt'a(t')\int^{t'}\dfrac{dt''}{a(t'')^3}+\dfrac{1}{3a}\omega^0_{ii}{\displaystyle \int dt' a(t')+\dfrac{{\cal C}}{a}}\,,\\ 
\label{ciit} 
c_i(t)&=&-{\displaystyle \int \dfrac{dt'}{3a(t')^3}\int^{t'} dt''a(t'')\sigma^0_{kki}+\int\dfrac{{\cal C}_i dt'}{a(t')^3}}\\ \nn
&&+\dfrac{1}{4}{\cal C}^0_{ikk}\int^t \dfrac{dt'}{a(t')^3}\int^{t'}dt''a(t'')\int^{t''}\dfrac{dt'''}{a(t''')^3}\,,
\eea
where we made the time dependence explicit by introducing the constant coefficients $  \cal C $, $  {\cal C}_{i} $, $  \omega^{0}_{ij} $, $  \tilde \omega_{ij} $, $  \sigma_{ijk}^{0} $ and $\tilde  \sigma_{ijk} $. 
The $  \e_{0} $ Taylor coefficients in \eqref{e0} are now simply read off from \eqref{Ai}, \eqref{Bij} and \eqref{Cijk}.  They are fixed by the choice of $  \e^{i} $ up to the integration constants $ {\cal A}^0_i  $ and $ {\cal B}^0_{ij}  $. So the final formula for $\epsilon_0$ is
\ba
\epsilon_0 &=&\dfrac{1}{3a}\tilde{\omega}_{ii}\int^t dt'a(t')\int^{t'}\dfrac{dt''}{a(t'')^3}+\dfrac{1}{3a}\omega^0_{ii}{\displaystyle \int dt' a(t')+\dfrac{{\cal C}}{a}}\\ \nn
&&+
\dfrac{1}{a}\left({\cal A}^0_i+\dfrac{1}{3}\sigma^0_{kki}\int dt'a(t')-{\cal C}_i-\dfrac{1}{4}{\cal C}^0_{ikk}\int^{t}dt'a(t')\int^{t'}\dfrac{dt''}{a(t'')^3}\right)x^i\\ \nn
&&+\dfrac{1}{2a}\left({\cal B}^0_{ij}-\tilde{\omega}_{ij}+\dfrac{1}{3}\mu^0_{kkij}\int^t  dt'a(t')-\dfrac{1}{4}{\cal D}^0_{ijkk}\int^t dt'a(t')\int^{t'}\dfrac{dt''}{a(t'')^3}\right)x^ix^j\\ \nn
&&+\dfrac{1}{3!a}({\cal C}^0_{ijk}-\tilde{\sigma}_{ijk})x^ix^jx^k+\dfrac{1}{4!a}\left({\cal D}^0_{ijkl}-\tilde{\mu}_{ijkl}\right)x^ix^jx^kx^l\,.
\ea

%%%%%%%%%%%%%%%%%%%%%%%%%%%%%%%%%%%%%%%%%%%%%%%

\section{Perfect fluid vs generic scalar field}\label{details}

The most convenient gauge for deriving an equation of motion for ${\cal R}_c$, is the comoving gauge. The scalar parts of the Einstein equations in this gauge are
\ba
\label{En00}
-\dfrac{2}{M_p^2}\delta \rho&=&12H^2N_1-\dfrac{3}{2a^2}\nabla^2 {\cal R}_c-12H\dot{{\cal R}}_c+\dfrac{3H}{a^2}\nabla^2 \psi-\dfrac{1}{a^2}\nabla^2 N_1-\dfrac{1}{a^2}\nabla^2 \dot{\psi}\,,\\ \nn
\dfrac{2}{M_p^2}\delta p&=&4H\dot{N}_1+4(H^2+\dfrac{2\ddot{a}}{a})N_1-\dfrac{1}{2a^2}\nabla^2 {\cal R}_c+2\ddot{{\cal R}}_c\\ 
\label{Enij}
&&\hspace{4cm}+\dfrac{H}{a^2}\nabla^2 \psi+6H\dot{{\cal R}}_c+\dfrac{1}{a^2}\nabla^2 N_1+\dfrac{1}{a^2}\nabla^2 \dot{\psi}\,, \\ 
N_1&=&-\dfrac{1}{2H}\dot{{\cal R}}_c\,,\\ 
0&=&2N_1-{\cal R}_c+2\dot{\psi}+2H\psi\,.
\ea
The key assumption that enables us to find a single equation for the curvature is that $\delta \rho$ and $\delta p$ are related by
\be
\label{prho}
\delta \rho=\dfrac{\dot{\rho}}{\dot{p}}\delta p\,.
\ee
This is true when there is an equation of state $\rho(p)$. Indeed, the above equality holds for any adiabatic perturbation and in presence of multiple perfect fluids. The reason is that for adiabatic modes we have
\be
\dfrac{\delta \rho}{\dot{\rho}}=\dfrac{\delta p}{\dot{p}}=\e_0\,.
\ee
Using \eqref{prho} one can remove $\delta \rho$ and $\delta p$ from a linear combination of \eqref{En00} and \eqref{Enij}. Then after some algebra, one can derive an equation, solely for ${\cal R}_c$ (neglecting the $\nabla^2 {\cal R}_c$ term):
\be
\ddot{{\cal R}}_c+\dfrac{\dot{f}}{f}\dot{{\cal R}}_c&=&0\,, \quad \text{with} \quad f(t)=a^3 \dfrac{\dot{H}}{H^2}\dfrac{\dot{\rho}}{\dot{p}}\,.
\ee
By means of the background equations, it is easy to see that the solution to this equation
\be
{\cal R}_c=\int dt' \dfrac{H^2}{a(t')^3 \dot{H}}\dfrac{\dot{p}}{\dot{\rho}}\,,
\ee
coincides \eqref{sol3} up to an integration constant. 
 
One may still wonder how does a scalar field cease to satisfy \eqref{prho}. The point is that, in comoving gauge and for a scalar field, $\e_0$ must vanish. So \eqref{prho}, as well as \eqref{En00} and \eqref{Enij} shrink into zero, up to leading order in momentum $q$. Actually, in contrast with a generic fluid, the $\delta \rho$ and $\delta p$ associated with a canonical scalar field, are not independent from metric perturbations. In the comoving gauge they are given by
\cite{WB} 
\be
\delta \rho=\delta p=-N_1 \dot{\phi}^2\,,
\ee
which is completely different than \eqref{prho}. For the perturbed Einstein equations to be consistent, one of the scalar equations must then be trivial. This is indeed the case as the momentum conservation equation becomes proportional to background dynamical equation of motion and hence trivial. As a result, the Bianchi identity implies that one of the scalar equations is redundant and the wholes set of equations is consistent. The resulting second order equation for ${\cal R}_c$ becomes
\be
{\cal R}_c''+\dfrac{2z'}{z}{\cal R}'_c-\dfrac{\nabla^2}{a^2}{\cal R}_c=0\,,
\ee
where $ z=a\dot{\phi}/H  $, and prime stands for derivative with respect to the conformal time. The answer to the above differential equation, neglecting the Laplacian, is
\be
{\cal R}_c=a_1+a_2 \int^t \dfrac{H(t')^2dt'}{\dot{H}(t')a^3}\,,
\ee
which clearly differs from \eqref{Newaddi}.

%%%%%%%%%%%%%%%%%%%%%%%%%%%%%%%%%%%%%%%%%%
 
\section{Viscous fluids}\label{app:viscous}

An imperfect fluid is described by the following energy-momentum tensor
\ba
\label{imperfect}
T_{\mu\nu}&=&(\rho+p)u_\mu u_\nu+p g_{\mu\nu}\\ \nn
&& -2\eta \left(\nabla_{(\nu}u_{\mu)}+u^\kappa \nabla_{\kappa}u_{(\mu}u_{\nu)}\right)\\ \nn
&&-(\zeta-\dfrac{2}{3}\eta)\nabla_\kappa u^\kappa \left(g_{\mu\nu}+u_\mu u_\nu\right)\,,
\ea
where $\zeta$ and $\eta$ are bulk and shear viscosities, respectively, and $u^\mu$ stands for the energy transport four-velocity. Consider a homogeneous and isotropic solution
\be 
\rho,p,\zeta,\eta=\text{depend only on time}\quad \text{and}\quad u^\mu=(1,0)\,.
\ee
The background energy-momentum tensor becomes
\be
\bar{T}_{00}=\bar{\rho}\quad,\quad \bar{T}_{ij}=(\bar{p}-3H\bar{\zeta})\delta_{ij}\,.
\ee
The first order perturbations of the energy-momentum tensor will be
\ba
\delta T_{00}&=&\delta\rho-\brho h_{00}\,,\\ \nn
\delta T_{0i}&=&-(\brho+\bp-3H\bzeta)\delta u_i+(\bp-3H\bzeta)h_{0i}\,,\\ \nn
\delta T_{ij}&=&(\bp-3H\bzeta)h_{ij}+a^2(\delta p-3H\delta \zeta)\delta_{ij}\,,\\ \nn
&&+\baeta \left[-2\partial_{(i}\delta u_{j)}+2\partial_{(i}h_{0j)}-a^2\partial_i \partial_j \dot{B}-2a^2 \partial_{(i}\dot{C}_{j)}-\dot{D}_{ij}\right]\,,\\ \nn
&&+\baeta a^2\delta_{ij}\left[\dfrac{1}{3}\nabla^2\dot{B}-\dfrac{2}{3a}\nabla^2F+\dfrac{2}{3a^2}\nabla^2 u_S\right]\,,\\ \nn
&&+a^2\bzeta \delta_{ij}\left[\dfrac{3}{2}HE-\dfrac{1}{2}\nabla^2\dot{B}+\dfrac{1}{a}\nabla^2 F-\dfrac{1}{a^2}\nabla^2 u_S-\dfrac{3}{2}\dot{A}\right]\,.
\ea
The energy density, the pressure and the velocity perturbations, according to their definitions in \cite{WB}, are given by
\ba
T_{00}&=&\rho^W \,,\\ \nn
T_{ij}&=& p^W\delta_{ij}\,,\\ \nn
\delta T_{00}&=&\delta \rho^W-\rho^W h_{00}\,,\\\nn
\delta T_{0i}&=&p^W h_{0i}-(\rho+p)^W \delta u_i^W\,,\\ \nn
\delta T_{ij}&=&p^W h_{ij}+a^2\left[\delta p\delta_{ij}+\partial_i\partial_j \pi^S+2\partial_{(i} \pi_{j)}^V+\pi^T_{ij}\right]\,.
\ea
For an imperfect fluid, these quantities are different than the ones in \eqref{imperfect}, so we used $W$ superscript to distinguish between them. The relationship among them turns out to be 
\ba
\rho^W &=& \brho\,,\\ \nn
p^W &=&\bp-3H\bzeta\,,\\ \nn
\delta u_i^W&=&\delta u_i\,,\\ \nn
\delta p^W&=&\delta p-3H\delta \zeta+\bzeta\left[\dfrac{3}{2}HE-\dfrac{1}{2}\nabla^2\dot{B}+\dfrac{1}{a}\nabla^2 F-\dfrac{1}{a^2}\nabla^2 u_S-\dfrac{3}{2}\dot{A}\right]+\\ \nn
&&\baeta\left[\dfrac{1}{3}\nabla^2\dot{B}-\dfrac{2}{3a}\nabla^2F+\dfrac{2}{3a^2}\nabla^2 u_S\right]\,,\\ \nn
\pi^S &=&\dfrac{1}{a^2}\eta\left(2aF-2\delta u_S-a^2\dot{B}\right)\,,\\ \nn
\pi_i^V &=&\dfrac{1}{a^2}\eta \left(aG_i-a^2\dot{C}_i-\delta u_i^V\right)\,,\\ \nn
\pi_{ij}^T &=&-\eta \dot{D}_{ij}\,.
\ea
%%%%%%%%%%%%%%%%%%%%%%%%%%%%%%%%%%%%%%%%%%%%

\section{Adiabatic Modes in Comoving Gauge} \label{app:comoving}

As pointed out earlier, the $\e^i$ as the spatial part of the residual diff in comoving gauge, satisfies the same equation as its Newtonian counterpart \eqref{epi2}. So expanding it up to quadratic order
\be
\label{eic}
\e^{i}=\dfrac{1}{a^2}\ep_i=\bc_i(t)+\bom_{il}(t)x^l+\dfrac{1}{2!}\bs_{ikl}(t)x^kx^l+\mathcal{O}(x^{3})\,,
\ee
we come up with the following conditions
\ba
\label{bsig}
\bs_{ijk}=\bs_{ikj}\quad \text{and}\quad \bs_{ijj}=-\dfrac{1}{3}\bs_{jji}\,,
\ea
where we indicate the Taylor coefficients with capital greek letters in order to avoid confusion with the Newtonian gauge results. 

The off-diagonal and spatial parts of the Einstein equations, separately for scalars and vectors are
\ba
\label{00c}
\partial_i \partial_j \left(N_1+\Rc+\dot{\psi}+H\psi\right) &=& 0\,,\\
\label{Nvcond} 
\partial_j\left(\dot{N}^{V}_i+H N^{V}_i\right) &=& 0 \,, \label{save1}
\ea
while \textit{$0i$} components read
\ba
\label{0ic}
\partial_{i}\left( H N_1-\dot{\Rc} \right)&=&0 \,,\\ 
\nabla^2 N_i^V &=& -4\dot{H}a^2\delta u^V_i \,,\label{save2}
\ea
Following the same line of logic as in Newtonian gauge, by demanding that the above equations hold, the time dependence in $\e^i$ coefficients become specified. The final solutions to the metric and matter fields as well as various matrices exploited to represent the adiabatic modes are summarized in tables \ref{symmmatric} and \ref{diffcom}. Here we briefly enumerate the adiabatic modes. 

%%%%%%%%%%%%%%%%%%%%%%%%%%%%%%%%%%%

\begin{table}
\begin{center}
  \begin{tabular}{|c| c |c | c | c|}
  \hline
  \rowcolor{gray}
   & equation &Constraints &No. elements \\ 
   \hline
   &&&\\
   $\bar{{\cal C}}_i$ &-&-& 3\\ 
   &&&\\
   $\beta^0_i$&-&-&3\\ 
   &&&\\
   $\bom^0_{ij}$ &-& symmetric & 6 \\
   &&&\\
   $\tilde{\bom}_{ij}$&-& - & 9\\ 
   &&&\\
   $\theta^0_{ij}$&\ref{divepF}& symmetric+$\left(\theta_{kk}=-\tilde{\bom}_{kk}\right)$& 5\\
   &&&\\
   $\bs^0_{ijk}$ &\ref{epi2}& sym. in j$\leftrightarrow$k + ($\bs^0_{jji}=-3\bs^0_{ijj}$) & 15\\ 
   &&&\\
   $\tilde{\bs}_{ijk}$ &\ref{epi2}& sym. in j$\leftrightarrow$k + $(\tilde{\bs}_{kki}=-3\tilde{\bs}_{ikk})$ & 15\\ 
   &&&\\
  $\lambda^0_{ijk}$& \ref{divepF}& symmetric+
  $\left(\lambda^0_{iik}=-\tilde{\bs}_{iik}\right)$& 7\\
  &&&\\
$\tilde{M}_{ijkl}$&\ref{epi2}& \text{sym in jkl}+$\left(\tilde{M}_{kkij}=-3\tilde{M}_{ijkk}\right)$ & 21\\ 
   &&&\\
   $\pi^0_{ijkl}$&\ref{divepF}& symmetric+$\left(\pi^0_{iikl}=-\tilde{M}_{iikl}\right)$ & 9\\
    \hline
  \end{tabular}
\end{center}
\caption{Matrices used in table \ref{diffcom} to parametrize the adiabatic modes in the comoving gauge. In the second column, we have specified the equation, if any, that constrains the corresponding matrix elements.\label{symmmatric}}
\end{table}
%%%%%%%%%%%%%%%%%%%%%%%%%%%%%%%%%%

\subsection*{Known adiabatic modes}\label{ssec:}

\begin{itemize}
\item \textbf{Weinberg's first scalar mode (${\cal O}(\vec{x}^{0})$)} This is the same mode as \refeq{1} in Newtonian gauge, for which the only non-zero coefficient is 
\be
\Omega_{ij}=\Omega^0_{ij}=\dfrac{1}{3}\lambda \delta_{ij}\,,
\ee
and so we find
\ba
\Rc &=& -\dfrac{1}{3}\lambda\,,\\ \nn
\psi &=&  \dfrac{\lambda}{3a}\int a(t') dt'\,,\\ \nn
\delta \rho&=& 0\,.
\ea
%%%%%%%%%%%%%%%%%%%%%%
\item \textbf{Vanishing Weinberg's second scalar mode, ${\cal O}(\vec{x}^{0})$}

Weinberg's second scalar mode in Newtonian gauge, \refeq{2}, completely disappears in comoving gauge. One way to see this is to consider the equivalent non-zero Taylor coefficient, namely $\alpha_0$:
\be
F(\bfx)=\alpha_0\,,
\ee
which induces
\ba
\psi&=&\dfrac{\alpha_0}{a}\,.
\ea
This spatially constant term does not produce any metric perturbations. A second way to see the same thing is to start with the mode in Newtonian gauge and perform the gauge transformation to comoving gauge, namely the transformation that cancels the scalar velocity perturbations 
\be
\delta u |_{\text{comov.}}=\delta u|_{\text{Newt.}}+\Delta \delta u=\delta u|_{\text{Newt.}}-\e_{0}\overset{!}{=}0\,.\label{2a}
\ee
But this gauge transformation was exactly the one that created the mode in Newtonian gauge and so in comoving gauge the adiabatic mode completely disappears. However, things change once one considers a finite momentum perturbation. To see this, consider the above time shift as a $q\to 0$ limit of a $q$ dependent temporal diff, i.e. $\epsilon_0(\vec{q})$. Applying it to the \eqref{2} solution in Newtonian gauge, the only non-vanishing components of the transformed metric perturbation would be $\psi_q$, given by
\be
\lim_{q\to 0}\psi(\vec{q})=-\epsilon_0=\dfrac{{\cal C}}{a}\,.
\ee
Now we ask this naive question: is this solution adiabatic in comoving gauge? The answer is yes and no! Up to ${\cal O}(\vec{x}^0)$ order this piece does not contribute to $g_{0i}$ component, simply because it is a constant. The local effect of $\psi_q$ start from ${\cal O}(\vec{x})$ as seen by the following expansion
\be
g_{0i}=\partial_i \psi=\partial_i \left[\psi_{\vec{q}}\exp(i\vec{q}.\vec{x})\right]=i\vec{q}.\vec{x}\psi_q+{\cal O}(\vec{x}^2)\,,
\ee
Strictly speaking, the long wavelength behavior of this Fourier mode is contained in the ${\cal O}(\vec{x})$ adiabatic mode that will be discussed later.  

%%%%%%%%%%%%%%%%%%%%%%%%%%%%%
\item \textbf{Weinberg's tensor mode, ${\cal O}(\vec{x}^{0})$} This is the same mode as \refeq{3} in Newtonian gauge and corresponds to the same choice of coefficients
\ba
\Omega_{ij}= \Omega^0_{ij} &,& \Omega^0_{kk}=0\,, \\ 
\gamma_{ij}&=&\Omega^0_{ij}\,.
\ea

%%%%%%%%%%%%%%%%%%%%%%%%%%
\item \textbf{Gradient scalars (SCT), ${\cal O}(\vec{x})$} This is the same mode as \refeq{4} in Newtonian gauge, arising from the following choice of parameters
\ba
\Sigma_{ijk}=\Sigma^0_{ijl}&=&\delta_{il}\bar{b}^0_j+\delta_{ij}\bar{b}^0_l-\delta_{jl}\bar{b}^0_i\,,\\ 
\displaystyle C_i(t)&=& -\bar{b}^0_i\int \dfrac{dt'}{a(t')^3}\int^{t'} dt''a(t'')\,,\\ 
\displaystyle \Rc&=&-\bar{b}^0_ix^i\,,\\ 
\psi&=&\Big(\dfrac{1}{a}\int dt' a(t')\Big)\bar{b}^0_i x^i\,.
\ea
%%%%%%%%%%%%
\item \textbf{Gradient tensors (SCT), ${\cal O}(\vec{x})$} This is the same mode as \refeq{5} in Newtonian gauge. As in that case, the remaining elements of $\Sigma^0_{ijk}$ produce 12 gradient modes of gravitons
\ba
\displaystyle C_i&=&-\dfrac{1}{3}\Sigma^0_{kki}\Big(\int \dfrac{dt'}{a(t')^3}\int^{t'} dt''a(t'')\Big)\,,\\ 
\gamma_{ij}&=&-(\Sigma^0_{ijl}+\Sigma^0_{jil}+2\Sigma^0_{lkk}\delta_{ij})x^l  \,,\\\nn
\psi &=&\dfrac{1}{3}\bs_{kkl}x^l\Big(\dfrac{1}{a}\int dt' a(t')\Big)\,,\\ \nn
\Rc&=&-\dfrac{1}{3}\sigma^0_{kkl}x^l\,.
\ea 
again the scalar part can be removed by subtracting a SCT with the following vector
\be
\bar{b}^0_i=-\dfrac{1}{3}\Sigma^0_{kki}\,.
\ee
\end{itemize}

%%%%%%%%%%%%%%%%%%%%%%%%%%%%%%%%%%%%%%%%
 
\subsection*{New adiabatic modes}\label{ssec:}

\paragraph{Pure modes}
\begin{itemize}
\item \textbf{Time dependent mode of ${\cal R}$, ${\cal O}(\vec{x}^0,\vec{x}^2)$ in scalars}

Choosing $\tilde{\bom}_{ij}=\dfrac{1}{3}\tilde{\bom}_{kk}\delta_{ij}$ corresponds to the following diff
\be
\e_0=\dfrac{1}{3\dot{H}a^3}\tilde{\bom}_{kk}\quad \text{and}\quad \e^i=\tilde{\bom}_{ij}x^j \int^t \dfrac{dt'}{a(t')^3}\,.
\ee
Using such a diff together with a non-zero $\theta^0_{ij}$, satisfying 
\be
\theta^0_{ij}=-\tilde{\bom}_{ij}\,,
\ee
yields the following adiabatic mode:
\ba
{\cal R}_c&=&\dfrac{1}{3}\tilde{\bom}_{kk}\left(\dfrac{H}{\dot{H}a^3}-\int^t \dfrac{dt'}{a(t')^3}\right)\\ \nn
\psi&=&\dfrac{1}{3}\tilde{\bom}_{kk}\left(\dfrac{1}{a}\int^{t}dt'a(t')\int^{t'}\dfrac{dt''}{a(t'')^3}-\dfrac{1}{\dot{H}a^3}\right)-\dfrac{1}{2a}\tilde{\bom}_{kk}\vec{x}^2\\ \nn
N_1&=&\dfrac{1}{3}\tilde{\bom}_{kk}\partial_t (\dfrac{1}{\dot{H}a^3})
\ea
This mode could also be obtained by coordinate transforming the \eqref{Newaddi}, from Newtonian to the comoving gauge. 
\item\textbf{${\cal O}(\vec{x})$ in scalars} 
A non-zero $\bar{\cal C}_i$ corresponds to a time-dependent translation and induces a divergenceless perturbation in $g_{0i}$. The latter can be either expressed as a scalar, or as a vector perturbation. The following choice yields the former 
\be
\beta^0_i=-\bar{{\cal C}}_i\,,
\ee
and induces the following pure gradient scalar mode
\be
\psi=-\dfrac{1}{a}\bar{{\cal C}}_ix^i \,.
\ee
This is the same mode as the one represented by \eqref{sgradd}, in Newtonian gauge. 
\item\textbf{${\cal O}(\vec{x}^0)$ in vectors}\\
Instead of the previous choice, putting $\beta^0_i=0$, gives a vector mode
\be
N^V_i=-\dfrac{\bar{{\cal C}_i}}{a}\,,
\ee
This is identical to the mode appeared in \eqref{vmode}. 
%%%%%%%%%%%%%%%%%%%%%%%%%%%%%%%
\item\textbf{${\cal O}(\vec{x})$ in vectors}\\ 
An antisymmetric $\tilde{\Omega}_{ij}$ is associated with a time-dependent rotation, and generates the perturbation below
\be
N_i^V=-\dfrac{1}{a}\tilde{\Omega}_{[ij]} x^j\,,
\ee
which is the counterpart of \eqref{grpv}, in Newtonian gauge. 
\end{itemize}
%%%%%%%%%%%%%%%%%%%%%%%%%%%%%%%

\paragraph{Mixed modes}
\begin{itemize}
\item\textbf{${\cal O}(\vec{x}^0)$ in tensor, ${\cal O}(\vec{x}^2)$ in scalars (or ${\cal O}(\vec{x}^0)$ in tensors, ${\cal O}(\vec{x})$ in vectors})\\
Consider a symmetric $\tilde{\Omega}_{ij}=\tilde{\Omega}_{(ij)}$ and put 
\ba
\theta^0_{ij}=-\tilde{\Omega}_{ij}\,,
\ea
this choice turns off vector gradients and a mixed tensor scalar mode remains
\ba
\label{o0t}
\displaystyle \gamma_{ij}&=& -2\tilde{\Omega}_{(ij)}\int \dfrac{dt'}{a(t')^3}\,,\\ \nn
\psi &=&-\dfrac{1}{2}\tilde{\Omega}_{(ij)}x^ix^j\,.
\ea
Alternatively, we could have put $\theta^0_{ij}=0$. This selection replaces the scalar perturbations by a gradient mode in vectors
\be 
\label{O1v}
N^V_i=-\tilde{\Omega}_{(ij)}x^j\,.
\ee
Above two modes, are identical to \eqref{tendec} and \eqref{mixv}. 

\item\textbf{${\cal O}(\vec{x})$ in tensors, ${\cal O}(\vec{x}^3)$ in scalars(or ${\cal O}(\vec{x})$ in tensors, ${\cal O}(\vec{x}^2)$ in vectors)}\\
The cousin of \eqref{gt3s} in comoving gauge may be derived by means of $\tilde{\Sigma}_{ijk}$ matrix, together with the following matrix
\be
\lambda_{ijk}=-\tilde{\Sigma}_{ijk}\,. 
\ee
This results in seven gradient modes in tensors, albeit mixed with ${\cal O}(\vec{x}^3)$ scalar perturbations
\ba
\displaystyle \gamma_{ij}&=& -2\tilde{\Sigma}_{ijk}x^k \Big(\int \dfrac{dt'}{a(t')^3}\Big)\,,\\ \nn
\psi &=& -\dfrac{1}{6a}\tilde{\Sigma}_{ijk}x^ix^jx^k\,.
\ea
Alternatively, we could also have removed scalars by selecting $\lambda_{ijk}=0$, at the expense of generating an ${\cal O}(\vec{x}^2)$ mode in vectors
\be
N^V_i=-\dfrac{1}{2}\tilde{\Sigma}_{ijk}x^jx^k \,.
\ee
\item \textbf{Vanishing of ${\cal O}(\vec{x})$ in vectors, ${\cal O}(\vec{x}^2)$ in scalars}\\
The $\theta^0_{ij}$ matrix itself gives a mixed scalar-vector mode, but scalars and vector in this case, cancel each other out, i.e. $N_i=0$.

One may wonder what happens to \eqref{o1vo2s} in Newtonian guage, once translated to the comoving gauge. The subtlety here is exactly the same as Weinberg's second adiabatic mode case. Namely, in the Newtonian gauge the diffeomorphism that creats \eqref{o1vo2s} is purely temporal. Consequently, it vanishes when we move to the comoving gauge.  

\item \textbf{${\cal O}(\vec{x}^0,\vec{x}^2)$ in vectors, ${\cal O}(\vec{x},\vec{x}^3)$ in scalars, ${\cal O}(\vec{x})$ in tensors}

Selecting a $\bs_{ijk}$, non-symmetric among $i$ and $j$, necessitates the presence of $\lambda^0_{ijk}$ due to the $\lambda^0_{iik}=-\tilde{\bs}_{iik}$ property. Subsequently, parallel to its Newtonian cousin \eqref{vmix}, there will be an ${\cal O}(\vec{x}^0)$ adiabatic mode in $\delta u^V_i$, iff $\tilde{\bs}_{ikk}\neq 0$. Such a solution is mixed both with scalars and tensors. The final answer would be 
\ba 
\delta u^V_i&=&\dfrac{1}{\dot{H}a^3}\tilde{\bs}_{ikk}\,,\\ \nn
{\cal R}_c&=&\tilde{\bs}_{ikk}x^i\left(-\dfrac{H}{\dot{H}a^3}+\int^t \dfrac{dt'}{a(t')^3}\right)\,,\\ \nn
N^V_i&=&-\dfrac{1}{2}\left(\tilde{\bs}_{ijk}+\lambda^0_{ijk}\right)x^jx^k\,,\\ \nn
\psi &=&\tilde{\bs}_{ikk}x^i\left(\dfrac{1}{\dot{H}a^3}-\dfrac{1}{a}\int^t dt'a(t')\int^{t'}\dfrac{dt''}{a(t'')^3}\right)+\dfrac{1}{6a}\lambda^0_{ijk}x^ix^jx^k\,,\\ \nn
N_1&=&-\tilde{\bs}_{ikk}x^i \partial_t (\dfrac{1}{\dot{H}a^3})\,,\\ \nn
\gamma_{ij}&=&-2\left(\tilde{\bs}_{lkk}\delta_{ij}+\tilde{\bs}_{(ij)l}\right)x^l\int^t\dfrac{dt'}{a(t')^3}\,.
\ea
\item \textbf{${\cal O}(\vec{x},\vec{x}^3)$ in vectors, ${\cal O}(\vec{x}^2,\vec{x}^4)$ in scalars, ${\cal O}(\vec{x}^2)$ in tensors}

The same gradient mode of $\delta u^V_i$ as in \eqref{vgmix}, appears in comoving gauge if we exploit a $\tilde{M}_{ijkl}$ matrix that is not symmetric between $i$ and $j$. Then $\pi^0_{kkij}=-\tilde{M}_{kkij}$ implies that $\pi^0_{ijkl}$ is nonzero. Finally, we find the following mixed adiabatic mode 
\ba
\delta u^V_i &=& \dfrac{1}{\dot{H}a^3}\tilde{M}_{(ij)kk}x^j\,,\\\nn
N^V_i&=&-\dfrac{1}{6a}(\tilde{M}_{ijkl}+\pi^0_{ijkl})x^jx^kx^l\,,\\ \nn
{\cal R}_c &=& \dfrac{1}{2}\tilde{M}_{(ij)kk}x^ix^j \left(-\dfrac{H}{\dot{H}a^3}+\int^t\dfrac{dt'}{a(t')^3}\right)\,,\\ \nn
\psi &=&\dfrac{1}{2}\tilde{M}_{(ij)kk}x^ix^j\left(\dfrac{1}{\dot{H}a^3}-\dfrac{1}{a}\int^{t}dt'a(t')\int^{t'}\dfrac{dt''}{a(t'')}\right)+\dfrac{1}{24a}\pi^0_{ijkl}x^ix^jx^kx^l\,,\\ \nn
N_1&=&-\dfrac{1}{2}\tilde{M}_{(ij)kk}x^ix^j\partial_t(\dfrac{1}{\dot{H}a^3})\,,\\\nn
\gamma_{ij}&=&-\left(\tilde{M}_{(lm)kk}\delta_{ij}+\tilde{M}_{(ij)lm}\right)\int^t\dfrac{dt'}{a(t')^3}x^lx^m\,.
\ea
\end{itemize}
\pagebreak
%%%%%%%%%%%%%%%%%%%%%%%%%%%%%%%%%%%% 
\begin{table}[H]
\begin{center}
\footnotesize
\begin{tabular}{|>{\columncolor[gray]{0.92}}c|c|}
\hlineB{3}
&\\
$\e_0$ & $\dfrac{1}{3\dot{H}a^3}\tilde{\bom}_{kk}-\dfrac{1}{\dot{H}a^3}\tilde{\bs}_{ikk}x^i-\dfrac{1}{\dot{H}a^3}\dfrac{1}{2}\tilde{M}_{(ij)kk}x^ix^j$\\ 
&\\
\hlineB{3}
&\\
$\epsilon_i$ & 
$
\begin{aligned}
&a^2\left(-{\displaystyle \int \dfrac{dt'}{3a(t')^3}\int^{t'} dt''a(t'')\Sigma^0_{kki}+\int\dfrac{\bar{{\cal C}}_i dt'}{a(t')^3}}\right)\\ 
&+a^2\left(\tilde{\Omega}_{ij}{\displaystyle\int \dfrac{dt'}{a(t')^3}+\Omega^0_{ij}}\right)x^j+\dfrac{1}{2}a^2\left(\tilde{\Sigma}_{ijk}{\displaystyle \int \dfrac{dt'}{a(t')^3}}+\Sigma^0_{ijk}\right)x^jx^k+\dfrac{1}{3!}a^2 \tilde{M}_{ijkl}\int^t \dfrac{dt'}{a(t')^3}x^jx^kx^l
\end{aligned}
$\\
&\\
\hlineB{3}
&\\ 
$F(\bx)$&$\alpha^0+\beta^0_ix^i+\dfrac{1}{2!}\theta^0_{ij}x^ix^j+\dfrac{1}{3!}\lambda_{ijk}^{0}x^ix^jx^k+\dfrac{1}{4!}\pi^0_{ijkl}x^ix^jx^kx^l$\\ 
&\\
\hlineB{3}
\hlineB{3}
\hlineB{3}
&\\
$\psi$& 
$
\begin{aligned}
&\dfrac{\alpha_0}{a}+\dfrac{1}{3}\bom_{kk}\left(\dfrac{1}{a}\displaystyle\int dt' a(t')\right)-\tilde{\bom}_{kk}\left(\dfrac{1}{3\dot{H}a^3}-\dfrac{1}{3a}\int^t dt'a(t')\int^{t'}\dfrac{dt''}{a(t'')^3}\right)\\ 
&+\dfrac{1}{3}\bs^0_{kki}x^i\left(\dfrac{1}{a}\displaystyle\int dt' a(t')\right)+\dfrac{\beta^0_i}{a}x^i+\tilde{\bs}_{ikk}x^i\left(\dfrac{1}{\dot{H}a^3}-\dfrac{1}{a}\int^t dt'a(t')\int^{t'}\dfrac{dt''}{a(t'')^3}\right)\\
&+\dfrac{1}{2}\tilde{M}_{(ij)kk}x^ix^j\left(\dfrac{1}{\dot{H}a^3}-\dfrac{1}{a}\int^{t}dt'a(t')\int^{t'}\dfrac{dt''}{a(t'')}\right)\\
&+\dfrac{1}{2a}\theta^0_{ij}x^ix^j+\dfrac{1}{6a}\lambda_{ijk}^{0}x^ix^jx^k+\dfrac{1}{24a}\pi^0_{ijkl}x^ix^jx^kx^l
\end{aligned}
$\\ 
&\\
\hlineB{3}
&\\
${\cal R}_c$ & $
\begin{aligned}
&\dfrac{1}{3}\Omega^0_{kk}-\tilde{\bom}_{kk}\left(-\dfrac{H}{3\dot{H}a^3}+\dfrac{1}{3}\int^t\dfrac{dt'}{a(t')^3}\right)\\ 
&+\dfrac{1}{3}\Sigma^0_{kki}x^i+\tilde{\bs}_{ikk}x^i\left(-\dfrac{H}{\dot{H}a^3}+\int^t\dfrac{dt'}{a(t')^3}\right)+\dfrac{1}{2}\tilde{M}_{(ij)kk}x^ix^j \left(-\dfrac{H}{\dot{H}a^3}+\int^t\dfrac{dt'}{a(t')^3}\right)
\end{aligned}$\\ 
&\\
\hlineB{3}
&\\
$\delta u^V_i$ & $\dfrac{1}{\dot{H}a^3}\left(\tilde{\bs}_{ikk}+\tilde{M}_{(ij)kk}x^j\right)$\\
&\\
\hlineB{3}
&\\
$N^V_i$& 
$-\dfrac{1}{a}\left[(\bar{{\cal C}}_i+\beta^0_i)+(\theta^0_{ij}+\tilde{\Omega}_{ij})x^j+\dfrac{1}{2}(\tilde{\Sigma}_{ijk}+\lambda_{ijk}^{0})x^jx^k+\dfrac{1}{6}(\tilde{M}_{ijkl}+\pi^0_{ijkl})x^jx^kx^l\right]$\\ 
&\\
\hlineB{3}
&\\
$N_1$ & $\left(\dfrac{1}{3}\tilde{\bom}_{kk}-\tilde{\bs}_{ikk}x^i-\dfrac{1}{2}\tilde{M}_{(ij)kk}x^ix^j\right)\partial_t(\dfrac{1}{\dot{H}a^3})$\\
&\\
\hlineB{3}
&\\
$\gamma_{ij}$& 
$
\begin{aligned}
&-2\Omega^0_{(ij)}+\dfrac{2}{3}\Omega^0_{kk}\delta_{ij}+\left(\dfrac{2}{3}\tilde{\Omega}_{kk}\delta_{ij}-2\tilde{\Omega}_{(ij)}\right)\displaystyle\int \dfrac{dt'}{a^3(t')}-2\tilde{M}_{(ij)kk}\int^t \dfrac{dt'}{a(t')^3}\int^{t'}dt'' a(t'')\int^{t''}\dfrac{dt'''}{a(t''')^3}\\
&-\left(\Sigma^0_{ijl}+\Sigma^0_{jil}+2\Sigma^0_{lkk}\delta_{ij}\right)x^l
-2\left(\tilde{\Sigma}_{(ij)l}+\tilde{\bs}_{lkk}\delta_{ij}\right)\int \dfrac{dt'}{a^3(t')} x^l\\
&-\left(\tilde{M}_{(lm)kk}\delta_{ij}+\tilde{M}_{(ij)lm}\right)\int^t\dfrac{dt'}{a(t')^3}x^lx^m
\end{aligned}
$\\ 
&\\
\hlineB{3}
\end{tabular}
\caption{Formulas for the residual diffeomorphisms of the comoving gauge, the integration constant $F(\bfx)$, introduced in \eqref{defF}, and the resulting adiabatic fields are summarized in this table.\label{diffcom}}
\end{center}
\end{table}
%%%%%%%%%%%%%%%%%%%%%%%%%%%%%%%

%%%%%%%%%%%%%%%%%%%%

\end{document}